# Mid-Infrared Spectroscopy of Uranus from the Spitzer Infrared Spectrometer: 2. Determination of the Mean Composition of the Upper Troposphere and Stratosphere

Glenn S. Orton[a], J. I. Moses[b], Leigh N. Fletcher[c], A. K. Mainzer[d], Dean Hines[e], H. B. Hammel[f], J. Martin-Torres[g], Martin Burgdorf[h], Cecile Merlet[c], Michael R. Line[i]

[a]MS 183-501, Jet Propulsion Laboratory, California Institute of Technology, 4800 Oak Grove Drive, Pasadena, California 91109, USA

[b]Space Science Institute, 4750 Walnut St., Suite 205, Boulder, CO 80301, USA

[c]Atmospheric, Oceanic & Planetary Physics, Clarendon Laboratory, University of Oxford, Parks Road, Oxford OX1 3PU, UK

[d]MS 321-535, Jet Propulsion Laboratory, California Institute of Technology, 4800 Oak Grove Drive, Pasadena, California 91109, USA

[e]Space Telescope Science Institute, 3700 San Martin Drive, Baltimore, MD 21218, USA

[f]Association of Universities for Research in Astronomy, 1212 New York Avenue NW, Suite 450, Washington, DC 20005, USA

[g]Instituto Andaluz de Ciencias de la Tierra (CSIC-INTA), Avda. De las Palmeras, 4, 18100, Armilla, Granada, Spain

[h]HE Space Operations, Flughafenallee 24, D-28199 Bremen, Germany

[i]Department of Astronomy and Astrophysics, University of California – Santa Cruz, Santa Cruz, California 95064, USA

Corresponding author: G. Orton[a]
glenn.orton@jpl.nasa.gov
818-354-2460

ABSTRACT

Mid-infrared spectral observations Uranus acquired with the Infrared Spectrometer (IRS) on the Spitzer Space Telescope are used to determine the abundances of $C_2H_2$, $C_2H_6$, $CH_3C_2H$, $C_4H_2$, $CO_2$, and tentatively $CH_3$ on Uranus at the time of the 2007 equinox. For vertically uniform eddy diffusion coefficients in the range 2200-2600 $cm^2$ $s^{-1}$, photochemical models that reproduce the observed methane emission also predict $C_2H_6$ profiles that compare well with emission in the 11.6-12.5 µm wavelength region, where the $\upsilon_9$ band of $C_2H_6$ is prominent. Our nominal model with a uniform eddy diffusion coefficient $K_{zz}$ = 2430 $cm^2$ $sec^{-1}$ and a $CH_4$ tropopause mole fraction of $1.6x10^{-5}$ provides a good fit to other hydrocarbon emission features, such as those of $C_2H_2$ and $C_4H_2$, but the model profile for $CH_3C_2H$ must be scaled by a factor of 0.43, suggesting that improvements are needed in the chemical reaction mechanism for $C_3H_x$ species. The nominal model is consistent with a $CH_3D/CH_4$ ratio of 3.0 ± 0.2 x $10^{-4}$. From the best-fit scaling of these photochemical-model profiles, we derive column abundances above the 10-mbar level of 4.5+01.1/-0.8 x $10^{19}$ molecule-$cm^{-2}$ for $CH_4$, 6.2 ± 1.0 x $10^{16}$ molecule-$cm^{-2}$ for $C_2H_2$ (with a value 24% higher from a different longitudinal sampling), 3.1 ± 0.3 x $10^{16}$ molecule-$cm^{-2}$ for $C_2H_6$, 8.6 ± 2.6 x $10^{13}$ molecule-$cm^{-2}$ for $CH_3C_2H$, 1.8 ± 0.3 x $10^{13}$ molecule-$cm^{-2}$ for $C_4H_2$, and 1.7 ± 0.4 x $10^{13}$ molecule-$cm^{-2}$ for $CO_2$ on Uranus. A model with $K_{zz}$ increasing with altitude fits the observed spectrum and requires $CH_4$ and $C_2H_6$ column abundances that are 54% and 45% higher than their respective values in the nominal model, but the other hydrocarbons and $CO_2$ are within 14% of their values in the nominal model. Systematic uncertainties arising from errors in the temperature profile are estimated very conservatively by assuming an unrealistic "alternative" temperature profile that is nonetheless consistent with the observations; for this profile the column abundance of $CH_4$ is over four times higher than in the nominal model, but the column abundances of the hydrocarbons and $CO_2$ differ from their value in the nominal model by less than 22%. The $CH_3D/CH_4$ ratio is the same in both the nominal model with its uniform $K_{zz}$ as in the vertically variable $K_{zz}$ model, and it is 10% lower with the "alternative" temperature profile than the nominal model. There is no compelling evidence for temporal variations in global-average hydrocarbon abundances over the decade between Infrared Space Observatory and Spitzer observations, but we cannot preclude a possible large increase in the $C_2H_2$ abundance since the Voyager era. Our results have implications with respect to the influx rate of exogenic oxygen species and the production rate of stratospheric hazes on Uranus, as well as the $C_4H_2$ vapor pressure over $C_4H_2$ ice at low temperatures.

## 1. Introduction

Determinations of atmospheric composition on Uranus are difficult because the low temperatures and low hydrocarbon column abundances make the planet extremely faint at mid-infrared wavelengths – a situation that differs from that of the other solar-system giant planets. Atmospheric dynamical mixing on Uranus is comparatively weak, perhaps as a result of the planet's small internal heat source, so that methane ($CH_4$) is not transported to very high stratospheric altitudes, and photochemical production of other disequilibrium hydrocarbons is correspondingly suppressed (e.g., Atreya et al. 1991). The cold atmospheric temperatures force methane to condense in the troposphere, and although some $CH_4$ vapor makes it up past the tropopause cold trap to help heat the stratosphere via absorption of sunlight in the near-infrared vibrational bands, the stratospheric methane abundance is very depleted in Uranus compared with the other outer planets.

The small amount of methane that does make it into the stratosphere is subject to photolysis by ultraviolet radiation, just as on the other giant planets, and the subsequent atmospheric photochemistry produces heavier hydrocarbons such as acetylene ($C_2H_2$), ethylene ($C_2H_4$), ethane ($C_2H_6$), methylacetylene (the $C_3H_4$ isomer hereafter designated $CH_3C_2H$), and diacetylene ($C_4H_2$), as well as methyl radicals ($CH_3$) and other reactive intermediates (see the photochemical models of Atreya and Ponthieu 1983, Summers and Strobel 1989, Bishop et al. 1990, Moses et al. 2005, and the *Spitzer Space Telescope* observations of Burgdorf et al. 2006). Oxygen-bearing constituents such as water ($H_2O$) and carbon dioxide ($CO_2$) can also be present in the stratosphere of Uranus from external sources such as cometary impacts, satellite debris, or Kuiper-belt grains (Feuchtgruber et al. 1997, 1999; Burgdorf et al. 2006). Abundance determinations for these hydrocarbon and oxygen species can help illuminate the details of the complex chemistry taking place in giant-planet atmospheres, as well as provide constraints on impact rates and the source of exogenic material on the giant planets.

However, detection of these species is difficult and requires sensitive instrumentation. Acetylene has previously been detected on Uranus from *Voyager* Ultraviolet Spectrometer (UVS) observations (Herbert et al. 1987, Yelle et al. 1989, Bishop et al. 1990) and Earth-based mid-infrared observations (Orton et al. 1987, Encrenaz et al. 1998). Ethane has been identified from the Voyager UVS occultations (Bishop et al. 1990) and tentatively from ground-based mid-infrared observations (Hammel et al. 2006). Stratospheric $H_2O$ has been reported from *Infrared Space Observatory* (ISO) observations (Feuchtgruber et al. 1997, 1999). Carbon monoxide, which is outside the spectral range of *Spitzer* spectra, has been detected in the infrared by Encrenaz et al. (2004) and in the submillimeter range by Cavalié et al. (2014). The *Spitzer Space Telescope* Infrared Spectrometer (IRS) is two orders of magnitude more sensitive than previous space-based instruments (Houck et al. 2004), allowing the first unambiguous infrared detections of $C_2H_6$, $CH_3C_2H$, $C_4H_2$, and $CO_2$ on Uranus, as well as providing abundance determinations for previously detected species like $CH_4$ and $C_2H_2$ (Burgdorf et al. 2006). Observations of these additional species supply important new constraints for photochemical models.

C$_2$H$_2$ observations of Uranus have hinted at possible increases in the acetylene abundance with time, with the apparent peak C$_2$H$_2$ mixing ratio increasing from ~10$^{-7}$ in the Voyager era (Bishop et al. 1990) to ~4 x 10$^{-7}$ in the ISO era (Encrenaz et al. 1998) to ~2 x 10$^{-6}$ in the Spitzer era (Burgdorf et al. 2006). The rotational pole of Uranus is nearly in its orbital plane, and the planet experiences unprecedented seasonal forcing as a result of this ~98° obliquity. Hydrocarbon production rates depend on solar insolation and thus will be seasonally variable, as may the stratospheric temperatures and circulation (e.g., Friedson and Ingersoll 1987, Conrath et al. 1990). To investigate the possibility of time variability in the global-average thermal structure or hydrocarbon abundances on Uranus due to the northern hemisphere slowly rotating into view after its long ~21-year winter, we have acquired new *Spitzer*/IRS disk-averaged spectra of Uranus in December 2007, just 10 days after the Uranus northern vernal equinox. In Orton et al. (2014), hereafter called Paper 1, we discuss (a) the acquisition, calibration, and reduction of these *Spitzer*/IRS spectra; (b) the derivation of the global-average thermal structure on Uranus from the analysis of the H$_2$ S(1), S(2), and S(3) quadrupole lines and the broad, continuum-like, collision-induced absorption features of H$_2$; and (c) the derivation of the tropospheric methane and helium abundance, again from the analysis of the broad continuum-induced absorption in the 9-20 μm region.

In the current paper, we focus on molecular features in the 7-17 μm region. The high signal-to-noise ratio of these recent *Spitzer*/IRS spectra allows us to derive much improved values for the abundances of various atmospheric gases over those described by Burgdorf et al. (2006). We use newly developed one-dimensional (1D) photochemical models to predict the vertical profiles of the molecular constituents in order to more accurately model the Spitzer spectra. The thermal structure and tropospheric methane profile used in the model are constrained from the observational analysis described in Paper 1. The resulting updated photochemical models provide notably improved fits to the Spitzer spectra compared with results from previous photochemical models developed to fit the *Voyager* and ISO data (e.g., Summers and Strobel 1989, Bishop et al. 1990, Moses et al. 2005); however, the apparent changes required by the new modeling do not conclusively suggest time variability in hydrocarbon abundances. We discuss the implications of our model-data comparisons with respect to molecular abundances, stratospheric photochemistry, vertical transport, and influx rates of exogenic oxygen-bearing species.

## 2. Observations

We observed Uranus over the course of 9.2 hours on 2007 December 16-17, obtaining spectra from four different modules of *Spitzer*/IRS: Short High (SH), Short Low (SL), Long High (LH) and Long Low (LL). Using all four modules, wavelengths in the 5-37 μm range were sampled with resolving powers R = λ/Δλ = 90-600 (see **Fig. 1**). For the work described in this paper, we used three spectral modules: SL1 (λ=7.46-14.05 μm, with λ/Δλ= 8.2667), LL2 (13.98-21.43 μm with λ/Δλ= 5.9048) and selected orders of SH (9.95-19.30 μm with λ/Δλ ~ 600; see Paper 1 for a more detailed description of these modules). The beam size (3.6-10 arcsec, depending on wavelength, see Paper 1) was larger than the 3.35 arcsec diameter of Uranus, so the planetary disk was not spatially

resolved — only globally averaged fluxes were obtained.  However, some information on longitudinal spatial variations could be obtained because the observations acquired with different modules were not simultaneous, so that each module sampled a different range of planetary longitudes as Uranus rotated beneath the slit (see Paper 1 for further details).  Apparent inconsistencies in the derived fluxes at wavelengths that overlap within the different modules suggest some longitudinal variability.  These inconsistencies complicate our current analysis; the derivation of $C_2H_2$ is particularly affected due to a 10% difference in flux between the LL2 and SL1 modules near ~14 microns, where acetylene emission is prominent.  The presence of similar discrepancies at wavelengths where other hydrocarbon emissions are present, combined with the lack of inconsistencies at wavelengths where the $H_2$ continuum dominates, suggests real longitudinal differences in stratospheric hydrocarbon emission rather than some kind of observational artifact such as a difference in the spillover of radiance from the disk of Uranus outside the 3.7-arcsec SL1 slit compared with the 10-arcsec LL2 slit (see Paper 1).  We therefore derive two different $C_2H_2$ abundances, depending on which module (LL2 or SL1) we are considering in the spectral fits, although mid-stratospheric temperature variations may also play a role in the apparent rotational variability.

>Figure 1

Complete details of the observations and the reduction/calibration of the spectra are presented in Paper 1.  We derive the total radiance uncertainty for the IRS spectra to be ~6% for the LL2 spectra.  For the SL1 spectra, an additional 4% uncertainty arises from corrections for overfilling of the slit, for a cumulative uncertainty of 7%.  The SH spectra were not photometrically trustworthy and needed to be adjusted to the amplitude and slope of lower-resolution SL1 and LL2 spectra at corresponding wavelengths.  In the worst case, the SH fitting adds another ~2% uncertainty to the SH spectra with little change in the total calibration uncertainties of 6-7%.

**3. Models**

**3.1. Radiative-transfer models.** We simulate the upwelling radiance from Uranus using a line-by-line radiative-transfer code, with disk-averaged radiances determined using a 10-stream trapezoidal quadrature in the cosine of the emission angle.  The atmospheric model for the radiative-transfer code extends from a maximum pressure of 10 bars to a minimum pressure of 1 nanobar with 10 grid points per decade of pressure.  For efficiency, most calculations were done with a minimum pressure of 100 nanobars, reserving the 1-nanobar lower pressure only for those cases in which emission from the $H_2$ S(2) or S(4) quadrupole lines needed to be considered amid $C_2H_6$ emission or $CH_4$ absorption, respectively.  Because clear-atmosphere models provide a good fit to the spectra, we ignore opacity effects from clouds and aerosols; a more detailed justification for ignoring particulate opacity is provided in Paper 1.  We also note here that particles at condensation levels that are consistent with the hydrocarbon profiles are too cold to contribute significantly to the upwelling radiance.

**3.1.1. Spectroscopic line parameters**. We determine $H_2$-$H_2$ collision-induced absorption (CIA) using the values tabulated by Orton et al. (2007a) from *ab initio* models, adding to those models the contributions of $(H_2)_2$ dimers (Schaefer and McKellar 1990). $H_2$-He and $H_2$-$CH_4$ CIA value are modeled using similar tables, created from programs emulating the *ab initio* calculations of Borysow et al. (1988) and Borysow and Frommhold (1986), respectively. Many of the spectroscopic parameters associated with discrete transitions that we cite below are now included in the most recent edition of the GEISA-2009 database (Jaquinet-Husson et al. 2011). Because we cannot determine the bulk composition from these spectra, we use the He/$H_2$ ratio of 15/85 by number determined by Conrath et al. (1987).

We use $CH_4$ and $CH_3D$ spectroscopic parameters as described by Brown et al. (2003), with line transitions, strengths, and quantum identifications for $CH_4$ taken from Ouardi et al. (1996) and for $CH_3D$ taken from Nikitin et al. (2000). $H_2$- and He-broadening widths were taken from Pine (1992) and Margolis (1996).

Spectroscopic parameters for $C_2H_6$ are taken from Vander Auwera et al. (2007), including their isotopic $^{13}C^{12}CH_6$ lines, and from di Lauro (2012). For both, we use the $H_2$-broadening widths of Halsey et al. (1988). Spectroscopic parameters for $C_2H_2$ and $CO_2$ are taken from the GEISA-2003 compilation (Jaquinet-Husson et al. 2003). For $C_2H_2$, the $H_2$/He-broadened line widths and their temperature dependence are taken from a parameterized fit to the results of Varanasi (1992). The spectroscopic parameters for $C_4H_2$ are taken from a list that was generated from the work of Jolly et al. (2010) and now published as a part of GEISA-2009 (Jaquinet-Husson et al. 2011). The spectroscopic parameters for $CH_3C_2H$ were taken from an unpublished list based on the work of Muller et al. (2002) that is now documented in the GEISA-2009 compilation. Spectroscopic line parameters for $CH_3$ are used as developed by Bézard et al. (1998), with a more recent update on the dipole moment by Stancu et al. (2005).

Measured values of the $H_2$/He-broadened line widths are not available for $CO_2$, $C_4H_2$, $CH_3C_2H$ and $CH_3$. Testing with pressure-broadened widths between 0.10 and 0.05 cm$^{-1}$, coupled with temperature dependences of $T^{-n}$ with n varying between 0.70 and 0.80, show negligible differences for the upwelling radiation at our spectral resolution because all the emission is confined to the stratosphere where line widths are dominated by Doppler broadening. We use nominal values of 0.07 cm$^{-1}$ for the width and n=0.75. In fact, testing on lines of more optically thick species also shows little sensitivity to the Lorentz width of lines or characterization of distant wings,.

As in Paper 1, spectroscopic parameters for the $H_2$ quadrupole lines are taken from Jennings et al. (1987) and Poll and Wolniewicz (1978), with broadening parameters given by Reuter and Sirota (1994), similar to the treatment of $H_2$ quadrupole lines in Fouchet et al. (2003).

**3.1.2. Non-Local thermodynamic equilibrium**. For all lines other than the $H_2$ quadrupole transitions (for which we assumed LTE conditions at all levels, following Trafton 1999), we use a non-LTE model that is an updated version of the methane model by Martín-Torres et al (1998) with a similar collisional scheme. The new version includes the calculation of the vibrational excitation of $CO_2$ and the hydrocarbons $CH_3$, $C_2H_6$ $C_2H_2$, $CH_3C_2H$, and $C_4H_2$. We assume that LTE is valid within the rotational levels of every vibrational-rotational band. Level populations are computed for stationary conditions, and stimulated emission and scattering are neglected, given the atmospheric temperatures and the wavelength of the transition considered for these species. The model includes exchanges of energy by vibrational-translational (V-T) and vibrational-vibrational (V-V) processes, as well as by radiative processes. The radiative and collisional productions and losses are combined in a number of statistical equilibrium equations for each vibrational level. Radiative processes include spontaneous emission, direct absorption of solar radiation, and the exchange of photons among the atmospheric layers. The latter is treated using the Full transfer by Optimized LINe-by-line (FUTBOLIN) radiative-transfer code (Martin-Torres and Mlynczak 2005, Kratz et al 2005).

Collisional rates are inferred with the same formalism used by Yelle (1991), assuming it is valid for $H_2$ as a collisional partner for all the hydrocarbons. Rates with this formalism have been compared with the few measurements of hydrocarbons with $H_2$, such as the work by Häger et al. (1979, 1980) and Chen et al. (1982). Rate values from Chadwick et al. (1993) have been used for $C_2H_2$ and extrapolated to the other hydrocarbon species. For methane, in addition to the collisional rates used by Martin-Torres et al. (1998), where only collisions with $N_2$, $O_2$, where taken into account, we also used relaxation rates for $CH_4$-$H_2$ from Hess and Moore (1976) for the V=3 levels and Menard-Bourcin et al. (2000, 2001, 2005) for the other states. For $CO_2$, we used the widely accepted values from Nebel et al. (1994). We also estimate uncertainties in the non-LTE source function to determine their contribution to the uncertainties of retrieved properties. Non-LTE source function uncertainties mainly arise from uncertainties in the collisional rates (some of which are unknown). These have been expressed as errors in the pressures corresponding to given values of the source function to Planck function ratio (J/B), and they are estimated as 25% for $CH_4$, $C_2H_2$, and $C_2H_6$, 50% for $CH_3$, $CH_3C_2H$, $C_4H_2$, and $C_3H_8$ and 15% for $CO_2$. The effect of these uncertainties will be included in the general discussion of cumulative uncertainties associated with the abundances of each of these constituents in Section 4. The vertical profiles of the J/B ratios for these molecules are illustrated in **Figure 2**.

>Figure 2.

**3.2. Photochemical models**. For the radiative-transfer model described above, the assumed vertical profiles of the hydrocarbons and other gaseous constituents can strongly affect the predicted molecular emission spectra. We initially adopted the published results from Moses et al. (2005) and then scaled their vertical profiles for the hydrocarbons and $CO_2$ by constant amounts to better fit the data, as needed. However, the eddy diffusion coefficients adopted in the Moses et al. (2005) models were designed to be consistent with the *Voyager*/UVS occultations (Bishop et al. 1990), and we found that the resulting hydrocarbon abundance profiles needed to be scaled by relatively large factors of ~2-8 to fit the data, with an even larger (three orders of magnitude) scaling factor required for $C_4H_2$. Burgdorf et al. (2006) also needed large scaling factors to explain their analysis. The size of the scaling factors needed to reconcile models and data, together with the improved atmospheric temperature structure described in Paper 1, prompted us to develop updated 1D photochemical models for Uranus.

Our new models use the Caltech/JPL KINETICS code (Allen et al. 1981) and are identical to those described in Moses et al. (2005), except that we now adopt the nominal thermal-structure discussed in Paper 1, alter the rate coefficients for a few of the reactions, and explore a wider range of eddy diffusion coefficient profiles. We use the "Model C" chemistry of Moses et al. (2005) — their favored reaction mechanism based on model-data comparisons for all the giant planets. However, we make some additional changes and updates to the mechanism: (a) we adopt the theoretically derived $k_0$ and $k_\infty$ values of Smith (2003) for the termolecular reaction $H + CH_3 + M \rightarrow CH_4 + M$; (b) we assume that $k_0 = 1.49 \times 10^{-29} \, T^{-1}$ cm$^6$ s$^{-1}$ ($T$ in K) for the reaction $H + C_2H_3 + M \rightarrow C_2H_4 + M$; (c) we assume that $k_0 = 7.5 \times 10^{-17} \, T^{-3} \, e^{-300/T}$ cm$^6$ s$^{-1}$ ($T$ in K) for the reaction $C_2H_3 + C_2H_5 + M \rightarrow C_4H_8 + M$; and (d) we use a reduced value of $4 \times 10^{-13}$ cm$^3$ s$^{-1}$ for the reaction $H + C_3H_5 \rightarrow CH_3 + C_2H_3$. Most of these changes are implemented to allow $C_2H_2$ recycling to be more efficient (see Section 5.5). The model considers ~70 hydrocarbon and oxygen species that are affected by ~500 chemical reactions. The potential exogenic oxygen-bearing molecules CO, $H_2O$, and $CO_2$ are introduced to the upper atmosphere using an assumed ablation profile for incoming icy grains (see Moses et al. 2000a) that follows the pressure dependence described in Moses (2001). The total incoming flux of these oxygen molecules is a free parameter adjusted to match observations. Condensation of $C_2H_2$, $C_2H_6$, $C_3H_8$, $C_4H_2$, $C_4H_{10}$, $C_6H_6$, $CO_2$, and $H_2O$ are included in the photochemical model, in the manner described in Moses et al. (2000b). Methane condensation is ignored in the models because the hydrocarbon photochemistry is insensitive to the tropospheric methane abundance and because the best-fit $CH_4$ profile is subsaturated in the stratosphere, which is not easily reproduced in the 1D model; therefore, we emphasize that the derived $CH_4$ profiles from the photochemical model are only relevant at altitudes above the tropopause.

## 4. Results

As **Fig. 1** illustrates, the *Spitzer*/IRS spectra of Uranus show evidence for emission from $CH_4$, $C_2H_2$, $C_2H_6$, $CH_3C_2H$, $C_4H_2$, $CO_2$, and possibly $CH_3$. In this section, we compare synthetic spectra from the photochemical-model results with the observed spectra, and we derive the corresponding best-fit molecular abundances (see **Table 1**). In Appendix A, we provide a table of temperatures and best-fit values for the volume mixing ratios of each of the constituents discussed below as a function of atmospheric pressure for our nominal model, with temperatures alone supplied for pressures less than $10^{-7}$ bars.

### 4.1. $CH_4$

Methane is the principal carbon-bearing compound in the atmospheres of the giant planets. On Uranus, the tropopause temperatures are so cold that the tropopause acts as an effective cold trap, forcing methane to condense in the upper troposphere to form clouds. Within and above the methane cloud region, $CH_4$ remains at saturated or sub-saturated values in the upper troposphere, and then diffuses upward into the stratosphere at the nearly constant mole fraction achieved at the tropopause cold trap, decreasing eventually at high altitudes due to molecular diffusion. The latitudinal variability of tropopause temperatures determined from the Voyager IRIS experiment (Conrath et al. 1998) does not indicate any temperatures that are so warm as to produce a "leak" in this cold trap, as might be the case for Neptune (Orton et al. 2007b).

**4.1.1. Stratospheric profile.** We derive the vertical distribution of $CH_4$ by fitting it (and its isotopologue $CH_3D$) to the 7.4-9.5 μm (1050-1340 cm$^{-1}$) region of the spectrum. The variations of stratospheric $CH_4$ influence the 7.4-7.9 μm (1260-1340 cm$^{-1}$) portion of the spectrum, where the methane band can be seen in emission. The observed emission allows strong constraints to be placed on the methane mole fraction in the ~0.2 mbar region of the upper stratosphere, which helps constrain the location of the methane homopause; however, the stratospheric $CH_4$ abundance above and below the 0.2-mbar region is not well constrained by the data. The necessary amount of 0.2-mbar methane can be supplied in the photochemical model from a family of possible solutions with different $CH_4$ mole fractions at the tropopause ($f_{CH4}$) and different stratospheric eddy diffusion coefficient ($K_{zz}$) profiles. In general, a lower value of $f_{CH4}$ requires stronger atmospheric mixing (i.e., higher values of $K_{zz}$) to carry up a sufficient amount of $CH_4$ to 0.2 mbar. By the same token, higher values of $f_{CH4}$ require lower $K_{zz}$ values in the stratosphere to reproduce the observed methane emission. The sensitivity of the methane results to $f_{CH4}$ and $K_{zz}$ (including $K_{zz}$ gradients) is discussed below.

For each assumed $K_{zz}$ value, the tropopause $CH_4$ mole fraction is varied until an optimal fit is found to the emission feature at the center of the $\nu_4$ band of $CH_4$ at 7.6-7.9 μm (1270-1320 cm$^{-1}$). We do not fit longer wavelengths in order to avoid the effects of the S(4) quadrupole of $H_2$, and we do not fit shorter wavelengths because of possible confusion with potential contributions from reflected sunlight. This fitting is also done iteratively with solutions for the $C_2H_2$ and $C_2H_6$ profiles, described below, because of the ~3% contribution from the 7.53-μm $\nu_4+\nu_5$ band of acetylene and the 7.25-μm $\nu_6$ band of ethane to the radiance of the 7.6-7.9 μm spectrum. (The poorer fit of the models at 1320-1330 cm$^{-1}$ might suggest that other emission features are also present, a possibility that could be explored with higher-spectral resolution data in the future.) The $CH_4$ mole fraction vertical profiles for a range of values for $K_{zz}$ that we test are shown in **Figure 3A**. The 7.4-10.0 μm (1000-1350 cm$^{-1}$) spectra corresponding to these values are shown in **Figure 3B**. The uncertainty of the fit for the value of the mole fraction is illustrated in **Figure 3C** for the nominal solution pair $K_{zz}$ = 2430 cm$^2$ sec$^{-1}$ with a $CH_4$ tropopause mole fraction, $f_{CH4}$, of 1.6 x 10$^{-5}$. **Figures 4A** and **B** show the values for the family of $K_{zz}$ vs. $f_{CH4}$ pairs that fit the data. Each identified $K_{zz}$ vs. $f_{CH4}$ pair provides a good fit to the methane emission feature, but the nominal solution pair is favored because it provides the best fit to the $C_2H_6$ spectrum without requiring any rescaling of the ethane abundance profile from the photochemical model (see section 4.2 below). The values and sources of uncertainty, including the non-LTE function for $CH_4$, are discussed below together with the $C_2H_6$ fit in Section 4.2. Note from **Fig. 3A** that the models that provide the best fit to the $CH_4$ emission all converge on a $CH_4$ mole fraction of 2.9 ± 0.4 x 10$^{-6}$ at 0.18 mbar, the pressure at the peak of the contribution function in the $\nu_4$ band, illustrating the strong sensitivity of the results to the CH$_4$ mole fraction at this pressure. The column abundance of methane and those of the other stratospheric constituents discussed in Section 5 are given in **Table 1**, along with other characteristics in **Table 2**.

>Figure 3
>Figure 4
>Table 1

We also investigate the quality of fit from a vertically sloped $K_{zz}$ profile (see Section 5.1) and explore changes in composition that are required if we assume the alternative temperature profile discussed in Paper 1 (see Section 5.2). In those sections, we also discuss the differences between the results of those models and the best-fit model with the nominal temperature profile and vertically uniform $K_{zz}$. For the remainder of this section the uncertainties we discuss pertain only to the model with the nominal temperature profile and vertically uniform $K_{zz}$.

**4.1.2. Tropospheric profile**. The variations of stratospheric CH$_4$ do not influence the 7.9-9.5 μm (1050-1260 cm$^{-1}$) portion of the spectrum, which is sensitive only to the distribution of CH$_4$ deeper than the tropopause and to the CH$_3$D/CH$_4$ ratio. We set the deep methane mole fraction to 3.2%, the value adopted by Karkoschka and Tomasko (2009, 2011) – see also Paper 1. This value is between the 2.3% mole fraction preferred by Lindal et al. (1987) and values as high as 4% discussed by Sromovsky et al. (2011), both analyzing Voyager Radio Subsystem (RSS) occultation data. The IRS spectra are not sensitive to this deep value independently of the CH$_4$ abundance just above the methane condensation level, as is also true of the Hubble Space Telescope STIS spectra analyzed by Karkoschka and Tomasko (2009, 2011) (E. Karkoschka, pers. comm.). We vary the distribution of CH$_4$ above the condensation level up through the temperature minimum using a sloped distribution that simulates the one used by Karkoschka and Tomasko (2011) in their Figure 10. The best-fit slope and its ±3σ values are illustrated in **Figure 5A**, with the corresponding spectra in **Figure 5B**. This slope is intermediate between those derived by Karkoschka and Tomasko (2011) for different regions of the planet using Hubble Space Telescope STIS spectra.

>Figure 5

**4.1.3. CH$_3$D / CH$_4$ Ratio**. Independent of the tropospheric CH$_4$ distribution, we also fit the CH$_3$D/CH$_4$ ratio, as illustrated in **Figure 5C**. For the best-fit case of $f_{CH4}$=1.6 x 10$^{-5}$ and $K_{zz}$ = 2430 cm$^2$ sec$^{-1}$, the CH$_3$D/CH$_4$ ratio is 3.0 +0.2/-0.1 x 10$^{-4}$, where the cited uncertainty arises from the fitting. Although there is overlap in the spectral regions influenced by the tropospheric CH$_4$ profile and the CH$_3$D/CH$_4$ ratio (**Figs. 5B, 5C**), our testing shows that the uncertainties in fitting the tropospheric CH$_4$ distribution do not propagate into significant uncertainties in the CH$_3$D/CH$_4$ ratio. Because this ratio is derived from the influence of CH$_3$D absorption on CH$_4$ absorption at the same atmospheric level, other potential sources of error are minimized. Adopting an extremely different temperature profile changes the best-fit ratio by only 0.3 x 10$^{-4}$ (see Section 5.2 and **Table 2**). Our derived ratio is consistent with recent measurements in the near infrared by Irwin et al. (2012) that yield a range for CH$_3$D/CH$_4$ of 2.4-3.8 x 10$^{-4}$. For a fractionation factor of 1.68±0.09 (Lecluse et al. 1996), our value for $CH_3D/CH_4 = 3.0$ +0.2/-0.1 x $10^{-4}$ in the nominal model implies a D/H ratio of (3.0 +0.2/-0.1 x $10^{-4}$ / 1.68 ± 0.09)/4 = 4.4 +0.6/-0.5 x $10^{-5}$ (the factor of ¼ signifies the fraction of H atoms in the molecule that a D atom is replacing). This value is consistent with the D/H ratio determined from the HD measurements made by the ISO Short-Wavelength Spectrometer (Feuchtgruber et al. 1999), whose 1σ range covers 4.0-9.0 x $10^{-5}$, and more recent and accurate measurements by the Herschel PACS experiment, yielding D/H = 4.4±0.4 x $10^{-5}$ (Feuchtgruber et al. 2013).

We note that the feature near 8.33 μm (1200 cm$^{-1}$) corresponds to the position of the $\nu_6 + \nu_8$ combination band of $C_4H_2$, although this cannot be taken as a positive identification. Stronger constraints on the $C_4H_2$ abundance are discussed below. Including or excluding this feature in the fit to the entire band had no significant affect on the either the tropospheric $CH_4$ profile or $CH_3D/CH_4$ results.

**4.2. $C_2H_6$**  As discussed in Paper 1, we first match $C_2H_6$ emission features near the 12.33 μm (814 cm$^{-1}$) $H_2$ S(2) quadrupole line in order to use that line to constrain the upper-atmospheric temperature structure accurately. Results for the temperature profile prove to be independent of the value for $K_{zz}$ that is chosen to fit the model, so long as the $C_2H_6$ features surrounding the $H_2$ S(2) line are reproduced accurately. We use vertical profiles of $C_2H_6$ that correspond to the family of $f_{CH4}$ - $K_{zz}$ models that provide the best fits to the $CH_4$ emission discussed in Section 4.1 (**Fig. 4A**). We focus on the fit to the SL1 observations, because the SH observations are scaled to the absolute radiance values of the SL1 spectrum.

The model with $K_{zz}$= 2430 cm$^2$ sec$^{-1}$ and $CH_4$ tropopause mole fraction of 1.6 x $10^{-5}$ provides the best simultaneous fit to both the $CH_4$ emission spectra (**Fig. 3B**) and to the $C_2H_6$ emission spectra (**Fig. 6A**). Adding the propagation of uncertainty from non-LTE effects for $CH_4$ along with fitting uncertainty, the 1σ range of values for $f_{CH4}$ is determined to be 1.4 x $10^{-5}$ to 1.9 x $10^{-5}$. The corresponding range for $K_{zz}$ is 2190 to 2600 cm$^2$ sec$^{-1}$, which is completely correlated with the values and uncertainties of $f_{CH4}$. We note that the vertical dropoff of the $C_2H_6$ mole fraction is so steep toward high altitudes that there were no significant differences between model spectra that assumed LTE and those for which the non-LTE source function (**Fig. 2**) was used. We also take into account the additional uncertainty associated with the absolute calibration (i.e. 7% for the SL1 spectrum, as described in Paper 1) for absolute values of the mole fraction and column abundances (**Table 1**) for both $CH_4$ and $C_2H_6$. For these molecules and those discussed below, we derive cumulative uncertainties in abundances by treating uncertainties from non-LTE and from absolute calibration as potential offsets and adding them linearly to the fitting uncertainties. For $C_2H_6$ and the remaining molecules discussed in this Section, the peak mole fraction and the atmospheric pressure to which it corresponds are given in **Table 2**. **Figure 6B** shows that the best model fit to the SL1 $C_2H_6$ spectrum provides a good fit to the higher-resolution SH spectrum, with no evidence for features of additional species in the residuals. As described in Paper 1, we needed to solve for the spectral resolution of the SH spectra in each order; in this case the best fit resolving power is R=736.

>Figure 6

**4.3. $C_2H_2$**. One of the brightest stratospheric emission features in the spectrum arises from acetylene. Our approach to matching these data and those of the constituents discussed below is to scale the abundance profiles derived from the same photochemical models used for the $CH_4$ and $C_2H_6$ analysis by a factor that its uniform with altitude. The resulting scaled vertical abundance profiles are used to generate model spectra as shown in **Figure 7A,** with corresponding abundance information provided in **Table 2**. The best fitting nominal model, with $K_{zz}$=2430 cm$^2$ sec$^{-1}$ and $CH_4$ tropopause mole fraction of 1.6 x 10$^{-5}$, requires an associated $C_2H_2$-profile scaling factor of 0.85±0.07 in order to fit the SL1 spectrum. This uncertainty includes the ±0.01 propagation of error through the 1σ uncertainty in the $f_{CH4}/K_{zz}$ solutions. Adding systematic uncertainties: the 7% uncertainty of the absolute calibration in this spectral region and the ±0.01 uncertainty in the scaling factor from the uncertainty of the $C_2H_2$ non-LTE source function (**Fig. 2**), the best fit to the scaling factor and its absolute value is 0.85±0.14.

>Figure 7

This spectral region containing acetylene emission is covered by both the LL2 and SL1 spectra, with an offset from the phase shift in rotational sampling discussed in Paper 1. The scaling factor required to match the LL2 spectrum is 1.05±0.14, 24% higher than the corresponding values for the SL1 spectrum. This cumulative uncertainty includes all potential sources of error, including the ±6% associated with absolute calibration of the LL2 spectrum. We include in **Table 1** only the results of the model whose scaling factor matches the SL1 spectrum, because the abundances of both $CH_4$ and $C_2H_6$ are constrained by the SL1 spectrum. The difference between the two reflects the fact that the LL2 and SL1 data sampled different longitudes and therefore may have different stratospheric temperatures or different abundances, which will be addressed in a subsequent study. For now, the difference thus represents one measure of the systematic uncertainty in characterizing the global mean. We note that the best fit to the SL1 data overall does not fit the peak of the band very well at the two points sampling the Q-branch near 13.70 μm (730 cm$^{-1}$). Simple modifications to the $C_2H_2$ profile do not work. Profiles that come closer to matching the Q-branch emission along with the P- and R-branches require an immediate drop of the VMR above the saturation level, which would only be consistent with acetylene being supplied from below, which is unphysical. Changes in the temperature structure, such as the alternative profile or making the thermosphere isothermal above the microbar level, scale the radiances of the P- and R-branches in the same way as the Q-branch. Although the observed flux corresponding to the peak Q-branch radiance should not have been in the saturated regime of the IRS detector, the difficulty of fitting this feature might indicate otherwise; it is the brightest feature in the averaged SL1 spectrum. It may also be possible that the formal measurement uncertainties associated with the feature underestimate its rotational variability. In any case, we do not expect the failure of the model to match the measured Q-branch to be the result of poorly understood chemical reaction mechanisms, as is the case with $CH_3C_2H$, as described in Section 4.4 below.

As expected, matching the high-resolution (SH) observations (**Fig. 7B**) yields similar results, because these spectra are scaled to the low-resolution observations. Because the Q-branch of this band clearly breaches the saturation limits of the detectors in the SH mode, we ignore these observations and do not include them in the central panel of this figure. Otherwise the same models matching the low-resolution data match the SH data extremely well, including a best fit to the spectral resolution of R=590-615, depending on the order. Residuals to this fit do not clearly indicate the presence of any additional molecular signatures amid the $C_2H_2$ emission.

>Figure 8

**4.4. $C_4H_2$ and $CH_3C_2H$**. **Figure 8** illustrates a portion of the SH order-13 spectrum that shows the spectroscopic features associated with the $\nu_8$ band of diacetylene ($C_4H_2$) and the $\nu_9$ band of methylacetylene ($CH_3C_2H$, also known as propyne and sometimes written more generically as $C_3H_4$). The spectral feature near 15.63 μm (640 cm$^{-1}$) is unidentified. The $C_4H_2$ and $CH_3C_2H$ features were originally detected and modeled by Burgdorf et al. (2006). Because of overlap between these bands, iterative fitting is done between them. The optimum resolving power matching the observed spectral features corresponds to R = 630. The best fit for $C_4H_2$ requires a scaling factor and fitting uncertainty of 1.16±0.07. The column density of $C_4H_2$ is limited by condensation in the lower stratosphere and thus depends strongly on the vapor pressure of the gas over $C_4H_2$ ice; a more thorough discussion of this issue is provided in Appendix B. Considering the ±0.12 uncertainty generated by the uncertainty of the favored $f_{CH4}/K_{zz}$ solution, and the ±0.08 uncertainty propagated by the uncertainty of the fit to the nearby feature of $CH_3C_2H$, plus systematic uncertainties: ±6% uncertainty associated with absolute calibration in this spectral region, the best-fit scaling and its cumulative uncertainty are 1.16 ± 0.22.

The best fit for $CH_3C_2H$ requires a scaling factor and fitting uncertainty of 0.43 ± 0.08. The fit in the short-wavelength/high-frequency side of the band (636-639 cm$^{-1}$) is not as good as the rest of the band; Burgdorf et al. (2006) had a similar problem. Model-data comparisons with higher-resolution ISO data for Jupiter and Saturn also seemed to have difficulty with fits to the entire methylacetylene band (e.g., Moses et al. 2000a, 2005), suggesting that there are problems with the line parameters for this band or with the predicted vertical slopes or other aspects of the model. Considering the additional ±0.07 uncertainty generated by the uncertainty of the $f_{CH4}/K_{zz}$ solution, and the ±0.02 uncertainty propagated by the uncertainty of the fit to the nearby feature of $CH_3C_2H$, plus the ±6% systematic uncertainty associated with absolute calibration, the best-fit scaling and its cumulative uncertainty are 0.43 ± 0.13. Uncertainties in the non-LTE function are negligible for both $C_4H_2$ and $CH_3C_2H$.

**4.5. $CH_3$**. **Figure 9** shows the SH order-12 spectrum near 16.50 μm (606 cm$^{-1}$) showing the most unambiguous feature in our spectrum that could be interpreted as the $\nu_2$ fundamental band of $CH_3$, which was sought unsuccessfully by Burgdorf et al. (2006). The best-fit scaling factor for the $CH_3$ profile was 1.5, with 1σ fitting uncertainty of ± 0.7. Without the uncertainties propagated by uncertainties in the non-LTE function and in the absolute calibration, the best fit corresponds to a marginal detection at ~2σ; thus we quote a 3σ upper limit of 2.1 for the scaling factor. This fit is consistent with the resolving power R=743 used to match the $H_2$ S(1) quadrupole line in the same SH order.

**4.6. $C_2H_4$**. The strongest band of $C_2H_4$ is the vibration-rotation $\nu_7$ band at 10.55 μm (948 cm$^{-1}$), which is in the range of the SL1 spectrum. We do not detect the band above the variance of the fit to the $H_2$ continuum there, which is consistent with its abundance profile predicted by the nominal model. Using model spectra, we determine that its abundance in our nominal photochemical model (equivalent to a column abundance of 1.1 x 10$^{13}$ molecule cm$^{-2}$ above the 10-mbar level) would need to be enhanced by a factor of about five in order to be detected at the 1σ level.

>Figure 10

**4.7. $CO_2$**. **Figure 10** shows an expanded portion of the SH order-14 spectrum with the Q-branch of the $\nu_2$ band of $CO_2$, identified by Burgdorf et al. (2006). We modify the rate of exogenic oxygen deposition in the model until a good fit to this feature is obtained. Our nominal model, which includes fluxes of CO, $CO_2$ and $H_2O$, suggests a $CO_2$ influx rate of 3 x 10$^3$ cm$^{-2}$ s$^{-1}$ (see Section 5.6). This fit is consistent with a spectral resolving power R=668. The resulting $CO_2$ mole fraction profile is shown in **Figure 11**, along with the best fitting profiles of the other molecules discussed above. The cumulative uncertainty from fitting (±9%), and propagation of non-LTE uncertainty (±8%) and absolute-calibration uncertainty (±6%) is ±23%. We note that this $CO_2$ band is sufficiently distant from line manifolds of $C_2H_2$ emission to avoid confusion, so the $CO_2$ fitting results do not depend on the $C_2H_2$ fits. The implications for the rate of oxygen deposition are discussed below in Section 5.6.

Appendix A provides in tabular form the values of temperatures and best-fit values for the volume mixing ratios of each of the constituents discussed above as a function of atmospheric pressure for the nominal thermal-structure model, with temperatures alone supplied for pressures less than 10$^{-7}$ bars.

**5. Discussion**

**5.1 Sloped $K_{zz}$ models**.

There is theoretical justification for vertically inhomogeneous profiles for the eddy diffusion coefficient, $K_{zz}$. Both breaking and non-breaking atmospheric waves propagating through the middle atmosphere are expected to result in effective $K_{zz}$ values that vary with the inverse square root of pressure in the stratosphere (Lindzen 1981). Although real atmospheres may not follow this relation exactly, there may well be a non-zero slope to the vertical profile of $K_{zz}$ in the stratosphere of Uranus. In fact, we assumed a uniform $K_{zz}$ profile only in order to test a large number of cases for the $CH_4$ and $C_2H_6$ distributions that are self-consistent with photochemical models and match the data. However, large $K_{zz}$ slopes can be ruled out because we would observe a homopause much higher in altitude than the shallow one that is implied by the model best fitting the hydrocarbon observations (as described in Section 4). In fact, the narrowness of the altitude region between the homopause and the condensation region for the hydrocarbon photochemical products prevents the gas-phase products from being used as effective tracers to help constrain the $K_{zz}$ profile (as is typical on the other giant planets), and it is therefore impossible to derive a $K_{zz}$ slope uniquely.

Nonetheless, we test to see whether a model assuming a sloped $K_{zz}$ profile would provide a better fit to the spectrum than the uniform one. Starting with an arbitrary low value for $K_{zz}$ at the tropopause (300 cm$^2$ s$^{-1}$), we test several sloped models, requiring a simultaneous fit to the CH$_4$ and C$_2$H$_6$ emission features in the spectrum, just as for the uniform-$K_{zz}$ models in **Fig. 4a** and **Fig. 6a**, respectively. **Figure 12** shows the best-fit $K_{zz}$ profile obtained with the sloped models, corresponding to $f_{CH4}$ = 2.0 x 10$^{-5}$ and $K_{zz}$ = 300(100/$P$)$^{-0.349}$ in the stratosphere, for pressure $P$ in mbar. Note that the sloped-$K_{zz}$ value matches the uniform-$K_{zz}$ value of 2430 cm$^2$ s$^{-1}$ at a pressure of 0.25 mbar, not much deeper than the pressure level where various CH$_4$ profiles converge (**Fig. 3A**), indicating a level at which the IRS spectra of CH$_4$ and C$_2$H$_6$ emission is particularly sensitive, and not much deeper than the methane homopause itself (located at 0.07 mbar for the nominal constant $K_{zz}$ model and at 0.04 mbar for the sloped $K_{zz}$ model). This best-fit sloped $K_{zz}$ model has somewhat different abundances than the uniform $K_{zz}$ model, with the tropopause mole fraction of CH$_4$, $f_{CH4}$ = 2.0 x 10$^{-5}$ in place of 1.6 x 10$^{-5}$, and abundances for other constituents that differ by as much as 45% (**Table 2**). The residuals to the spectral fit are not substantially different than for the uniform $K_{zz}$ model. From this we deduce that any number of sloped-$K_{zz}$ models for various $f_{CH4}$ and $K_{zz}$ profiles could be developed that fit the Spitzer data equally well.

>Figure 12
>Table 2

**5.2 Alternative temperature profiles.** An alternative temperature profile that is described in detail in Paper 1 models the upper stratosphere as isothermal between 1 µbar and 0.1 mbar (**Fig. 12**). This profile provides nearly the same quality of fit to the IRS temperature-constraining data. On the other hand, we do not consider it to be a realistic profile because it is inconsistent with a profile derived from Voyager UVS occultation data (e.g. Herbert et al. 1987) and requires an unphysical abrupt temperature gradient at the base of the thermosphere (**Fig. 12**). Nonetheless, we consider this temperature profile to be useful in evaluating the extremes that systematic uncertainties of the temperature structure have on derived values for the abundance profiles of atmospheric constituents. We follow the same fitting procedure as for the nominal profile, seeking a consistent photochemical model that fits both the CH$_4$ and C$_2$H$_6$ emission features.

For this temperature profile, we derive $K_{zz}$ = 590 cm$^2$ s$^{-1}$, which corresponds to a molar fraction of CH$_4$ at the tropopause $f_{CH4}$ = 7.0 x 10$^{-5}$. In fact, better-fitting models are possible for even higher values of $f_{CH4}$, but 7.0 x 10$^{-5}$ represents a fully saturated abundance at the tropopause, i.e. 100% relative humidity. Just as in the case for the uniform- and sloped-$K_{zz}$ models for the nominal temperature profile, we scale profiles of the other hydrocarbons and CO$_2$ to match the emission features in the IRS spectra. The column abundances corresponding to these values are given in **Table 2**. Analysis of Hubble Space Telescope STIS spectra of Uranus by Karkoschka and Tomasko (2009, 2011) and the radio-occultation results of Lindal et al. (1987) and Sromovsky et al.

(2011) do not indicate tropopause molar fractions of $CH_4$ anywhere near 100% humidity, and so this model is ultimately unrealistic. Nonetheless, pursuing differences between abundances derived on the basis of this alternative profile and the nominal profile can be taken as an extremely conservative approach to assessing the uncertainties in derived abundances propagated by temperature uncertainties. **Table 2** shows that the biggest difference is a factor of over ~4 between the $CH_4$ column abundances above the 100-mbar level in the two models, largely due to the very different mole fractions of $CH_4$ at the tropopause. The $C_2H_6$ column abundance in the alternative model is 54% higher than in the nominal model, and the $3\sigma$ $CH_3$ column-abundance upper limit is a factor of two higher than in the nominal model. Differences in column abundances for the other stratospheric molecules are less than 23%. In the troposphere, the $CH_4$ distribution is within the uncertainties associated with the nominal model fit, and the $CH_3D/CH_4$ ratio differs by only $0.3 \times 10^{-4}$, compared with the $1\sigma$ uncertainty of $\pm 0.2 \times 10^{-4}$ associated with the nominal model fit. Thus, based on this very conservative assessment of errors derived from uncertainties in temperature structure, the derived abundances and profiles are robust to within ~25% for nearly all constituents, with $CH_4$ and $C_2H_6$ being major exceptions. An additional uncertainty on the order of ~25% can be assumed, as well, based on the longitudinal variability of $C_2H_2$ sampled by the SL1 *vs.* LL2 spectra (**Figure. 7A**).

Using the radio-occultation profiles of Lindal et al. (1987) or Sromovsky et al. (2011) makes a negligible change in our retrieved stratospheric hydrocarbon and $CO_2$ vertical distributions, because we smoothed through those temperature structures in the stratosphere (see Figure 9 of Paper 1) and subsequently modified them to match the IRS spectra. On the other hand, the radio-occultation profiles of Lindal et al. (1987) and Sromovsky et al. (2011) differ from our standard temperature profile in the troposphere, as discussed in Paper 1. To varying degrees, these temperature profiles are colder than our standard profile in the region where $CH_4$ absorption is taking place. The primary difficulty with these models is in matching the observed spectrum that is also controlled by $H_2$ CIA absorption in the 1050-1090 cm$^{-1}$ (9.2-9.5 μm) region. The closest matches of the radio-occultation temperature profiles to this region are Model F of Lindal et al. (1987) (see Figure 17 of Paper 1) and Model F1 of Sromovsky et al. (2011) warmed uniformly by 1K (see Figure 18 of Paper 1), for both of which the $CH_4$ VMR at depth is ~4%. Assuming a vertical distribution model for tropospheric $CH_4$ that is similar to those shown in **Fig. 5A**, both models are several standard deviations away from the measured spectrum at 1050-1090 cm$^{-1}$. The higher wavenumbers (shorter wavelengths) can be fit for the Lindal et al. Model F for a $CH_4$ VMR profile that is nearly the same as our best fit and for the Sromovsky et al Model F'+1K for a profile that reaches the 3.2% VMR at 2.5 bars, not far from the Karkoschka and Tomasko (2011) model for a latitude of 45°S. For these cases, the retrieved $CH_3D/CH_4$ ratio is close to $2.6 \times 10^{-4}$ in place of $3.0 \times 10^{-4}$.

**5.3 Implications for trace-constituent chemistry and mixing.**

A key criterion to test the validity of models for stratospheric mixing is the consistency of the associated abundance profiles for $C_2H_6$ and $C_2H_2$ with measurements. In fact, for the case of $C_2H_6$, we have used this consistency test to establish our favored nominal model. Taking only the fitting uncertainties into account, we find that models with vertically uniform diffusion coefficients in the range $K_{zz}$ = 2200-2600 cm$^2$ sec$^{-1}$ have $C_2H_6$ profiles with scaling factors that are within 1$\sigma$ of 1.00. We are not able to narrow these bounds further using a similar consistency test for $C_2H_2$. The concordance of the $C_4H_2$ scaling factors with the $C_2H_2$ and $C_2H_6$ results, along with the dependence of the $C_4H_2$ fit to the wings of the $CH_3C_2H$ feature (**Fig. 5**), prevents us from using the $C_4H_2$ fits to constrain $K_{zz}$ further. (The disagreement of the $CH_3C_2H$ scalings with this otherwise self-consistent model will be discussed in Section 5.7.)

**Figure 8** illustrates the vertical distributions of the various atmospheric constituents that resulted from our photochemical model. The *Spitzer*-derived abundances discussed above refine the early work of Burgdorf et al. (2006) and tentatively identify $CH_3$ in the atmosphere of Uranus with an abundance that is much lower than in the other outer planets, e.g. three orders of magnitude lower than detected in Neptune (Bézard et al, 1999).

We derive low column abundances for Uranus' hydrocarbons, in comparison with the other giant planets (e.g., Bishop et al. 1995, Moses et al. 2004, Fouchet et al. 2009). This result is consistent with the hydrocarbons being confined to atmospheric pressures greater than ~0.1 mbar on Uranus, as a result of the slow rate of vertical transport in the stratosphere and a resulting low-altitude $CH_4$ homopause (e.g., Herbert et al. 1987; Yelle et al. 1987, 1989; Summers and Strobel 1989; Bishop et al. 1990). The sluggish vertical mixing, in turn, may be related to Uranus's low internal heat flux (e.g., Atreya 1986; Conrath et al. 1991). The Voyager ultraviolet spectrometer (UVS) observations (e.g., Broadfoot et al. 1986; Herbert et al. 1987; Yelle et al. 1987, 1989; Bishop et al. 1990) first made it clear that $K_{zz}$ was comparatively small on Uranus, at least at summer southern solstice at the time of the Voyager encounters. Our Spitzer observations confirm that stratospheric $K_{zz}$ values are still low on Uranus in 2007, despite several other observations that suggest that stronger tropospheric convective activity or other dynamical changes on Uranus may have occurred from the 1986 Voyager encounter to at least the 2007 equinox over a wide range of latitudes (e.g., Karkoschka 2001; Rages et al. 2004; Klein and Hofstadter 2006; Hammel and Lockwood 2007; Norwood and Chanover 2009; Irwin et al. 2009, 2011, 2012; Sromovsky et al. 2009, 2012a, 2012b, 2012c)

The original Voyager UVS observations further suggested that the strength of atmospheric mixing on Uranus could be substantially different at high latitudes as compared with low latitudes. For example, the solar far-ultraviolet reflection data analyzed by Yelle et al. (1989) for high-latitude regions implies a $K_{zz}$ value as low as 50 cm$^2$ s$^{-1}$ (see also Summers and Strobel 1989). The Voyager UVS solar occultation at near-equatorial latitudes, on the other hand, suggests larger $K_{zz}$'s of a few 100 to 10$^4$ cm$^2$ s$^{-1}$ at the methane homopause (Herbert et al. 1987, Bishop et al. 1990; Atreya et al. 1991; see also Summers and Strobel 1989) — values that are similar to what we derive from our more recent global-average Spitzer observations. Note that the lower latitudes weight our global observations at equinox more heavily, so the apparent similarity with low-latitude Voyager values is not too surprising. The low-latitude values for $K_{zz}$ derived from Voyager are sufficiently large that methane is carried to altitudes far enough above the $C_2H_x$ hydrocarbon condensation regions in the lower stratosphere (~0.1-20 mbar) that measurable column abundances of heavy hydrocarbons can be produced from $CH_4$ photolysis. That is not the case for the lower value of $K_{zz}$ = 50 cm$^2$ s$^{-1}$ derived near the polar regions, where the column abundance of the heavier hydrocarbons is significantly reduced (e.g., Yelle et al. 1989, Summers and Strobel 1989, Moses et al. 2005) such that hydrocarbon emission might be unobservable, even by Spitzer, if the high-latitude regions were representative of the planet as a whole.

The apparent high- vs. low-latitude differences in $K_{zz}$ and the methane homopause pressure level could result from stratospheric circulation (e.g. Yelle et al. 1989, Summers and Strobel 1989, Flasar et al. 1987, Friedson and Ingersoll 1987, Conrath et al. 1990), making the development of a 1D "global-average" atmosphere difficult, as both meridional and vertical gradients in hydrocarbon abundances could be present on Uranus (see also Karkoschka and Tomasko 2009). However, as we show in **Fig. 11** and discuss further in Section 5.8, global-average infrared observations from ground-based telescopes (Orton et al. 1987, 1990) and the ISO spacecraft (Encrenaz et al. 1998), as well as the *Spitzer* data presented here and in Burgdorf et al. (2006), appear reasonably consistent with each other in terms of the inferred $C_2H_2$ and methane mole fractions and upper limits; those derived abundances also compare qualitatively with those derived from the Voyager UVS near-equatorial solar occultation results (Herbert et al. 1987, Bishop et al. 1990). Conditions at the low-latitude UVS solar-occultation site may then be considered more-or-less representative of the bulk of the planet, and one-dimensional photochemistry/transport models are still useful. However, more detailed comparisons of our Spitzer-based models with the Voyager UVS occultation light curves will be required to determine if there is any evidence for true increases in the $C_2H_2$ abundance (and hence $K_{zz}$) with time on Uranus (see Section 5.8 below).

**5.4. Dependence of the $C_4H_2$ abundance on its vapor pressure over $C_4H_2$ ice.** The column abundances of complex hydrocarbons on Uranus are very sensitive to their saturation vapor pressures, as virtually all the stable hydrocarbon photochemical products will condense in the lower stratosphere. Accurately reproducing the Spitzer observations therefore requires accurate knowledge of the condensable hydrocarbon vapor pressures at low temperatures. The vapor pressures of $C_2H_x$ hydrocarbons have been well studied, but there are few or no measurements of the $C_3H_x$ and $C_4H_x$ vapor pressures at the very low temperatures relevant to Uranus. The more refractory the molecule is, the less likely we are to have information on its vapor pressure, as measuring its vapor pressure at the low temperatures relevant to Uranus's atmosphere poses experimental difficulties. The situation is particularly acute for $C_4H_2$. From our model-data comparisons, we find that photochemical models that use previously derived expressions for the $C_4H_2$ vapor pressure over $C_4H_2$ ice result in diacetylene abundances that are orders of magnitude too small to explain the Spitzer $C_4H_2$ emission. In Appendix B, we discuss the reasons and make new recommendations for the vapor-pressure expression of $C_4H_2$ vapor over $C_4H_2$ ice.

The resulting $C_4H_2$ profile from our nominal model needs to be scaled by only 1.16 to provide the best fit to the Spitzer data, suggesting that our adopted expression is reasonable. It is further worth noting that our models indicate that $C_4H_2$ supersaturations can exist within the condensation region because the $C_4H_2$ photochemical production outpaces the condensation for realistic descriptions of the condensation process (see also Summers and Strobel 1989, Pollack et al. 1987; Rages et al. 1990; Moses et al. 1995, 2000b).

**5.5. $C_2H_2$ and $C_2H_6$ abundance and implications.** One important way in which Uranus differs from Jupiter, Saturn, and even Neptune is that its stratospheric vertical transport is much weaker; our results near equinox verify this conclusion from Voyager observations near solstice. This sluggish atmospheric mixing ensures that the methane photolysis rate will peak at relatively deep pressures on Uranus, putting the resulting hydrocarbon photochemistry in a different physical regime than on the other giant planets. The photolysis at high pressures, combined with the narrow altitude range in which photochemical production is available before condensation begins, leads to (1) a difference in the total production rate of complex hydrocarbons, (2) a difference in the relative efficiency of methane recycling, and (3) a difference in the relative abundances of $C_2H_2$ and $C_2H_6$ on Uranus as compared to the other giant planets. For example, the $C_2H_6/C_2H_2$ ratio is ~11 at 1 mbar on Jupiter, ~10 at 0.5 mbar on Saturn, and ~25 at 0.2 mbar on Neptune (Moses et al. 2005), whereas our Spitzer analysis suggests the $C_2H_6/C_2H_2$ ratio is less than unity on Uranus, amounting to ~0.5 at 0.2 mbar. We now provide detailed chemical explanations for this behavior.

**5.5.1 Competing reactions**. One driving factor behind all three of these consequences is that the termolecular reaction (A): $H + CH_3 + M \rightarrow CH_4 + M$ (where M corresponds to any other atmospheric molecule) is much less efficient at low pressures than the termolecular recombination reaction (B): $CH_3 + CH_3 + M \rightarrow C_2H_6 + M$, due to the shorter lifetime of the excited intermediate $CH_4^*$ complex and greater likelihood that it will break apart to reform into $CH_3 + H$. The heavier $C_2H_6^*$ intermediate complex more efficiently redistributes the excess collisional energy through its vibrational modes. As a result, the low-pressure-limiting rate coefficient for reaction A is a few orders of magnitude smaller than that for reaction B. On Jupiter, Saturn, and Neptune, methane photolysis occurs at low pressures (high altitudes) due to the more vigorous atmospheric mixing, and the reaction $CH_3 + CH_3 + M$ is fast enough to compete with $H + CH_3 + M$ as a loss process for $CH_3$ despite the relatively large H abundances in the photolysis region. The ethane production that occurs through this mechanism is therefore efficient and dominates the total column-integrated production of $C_2H_6$ – and in fact is the most effective pathway for converting methane to $C_2H_x$ hydrocarbons throughout the stratospheres of Jupiter, Saturn, and Neptune (e.g., Moses et al. 2000a, Moses et al. 2005). In contrast, because $CH_4$ is photolyzed at relatively high pressures on Uranus, reaction A becomes much more efficient and dominates the total column loss rate of $CH_3$. The total column production rate of ethane is therefore smaller on Uranus than the other giant planets, methane is more efficiently recycled, and the column production rate of other complex hydrocarbons that have ethane as a possible source (including $C_2H_2$) is also reduced. The relatively high-pressure photolysis location also reduces the primary production mechanisms for unsaturated hydrocarbons such as $C_2H_2$ and $C_2H_4$ because reactions like $CH + H_2 + M \rightarrow CH_3 + M$ then dominate over $CH + CH_4 \rightarrow C_2H_4 + H$.

Note that because $CH_4$ photolysis is photon-limited rather than methane-limited on Uranus and the other giant planets, the lower methane abundance in the stratosphere of Uranus is not a factor in explaining the smaller overall complex hydrocarbon production; in fact, heliocentric distance is the main distinguishing factor in terms of the total $CH_4$ column-integrated photolysis rate on the giant planets. The column-integrated methane photolysis rate is comparable on Uranus in relation to the other giant planets (e.g., higher than on Neptune and only a factor of ~0.25 smaller than on Saturn), but the efficiency of methane recycling, combined with the small altitude separation between the peak hydrocarbon production regions and their condensation levels, makes complex hydrocarbon formation very inefficient on Uranus.

**5.5.2 Ethane recycling**. The dominant factor that affects the relative abundances of $C_2H_2$ and $C_2H_6$ is the efficiency of their production and loss processes within the narrow altitude region available for photochemistry. On Jupiter, Saturn, and Neptune, the photochemically active region extends across many pressure levels, and different production, loss, and recycling mechanisms operate at different pressures. In the low-pressure formation region on Jupiter, Saturn, and Uranus, the production rate for ethane exceeds its destruction rate, and although recycling is not prevalent, the loss rates cannot keep pace with production, and the $C_2H_6$ formed at high altitudes can flow downward to build up a large column abundance. At the higher pressures relevant to $CH_4$ photolysis on Uranus, ethane photolysis loss rates more evenly match the production rates, so that the net $C_2H_6$ production rate is reduced. Acetylene production and loss rates tend to be more balanced on all the giant planets, with production rates that exceed loss rates at low pressures and the reverse being true at high pressures; however, $C_2H_2$ can be relatively efficiently recycled in the middle and lower stratospheres because its photolysis products tend to reproduce the $C_2H_2$ quickly (e.g., Moses et al. 2000a, 2005). A key point is that at large enough pressures in the stratosphere (which would be within the $C_2H_2$ condensation region on Uranus), acetylene can be effectively converted to ethane through sequential addition of atomic hydrogen. This mechanism is also responsible for ethane recycling on the giant planets: $C_2H_6$ photolysis predominantly leads to the production of $C_2H_2$, and the most important recycling mechanism for $C_2H_6$ from a column-integrated standpoint then becomes the following series of reactions:

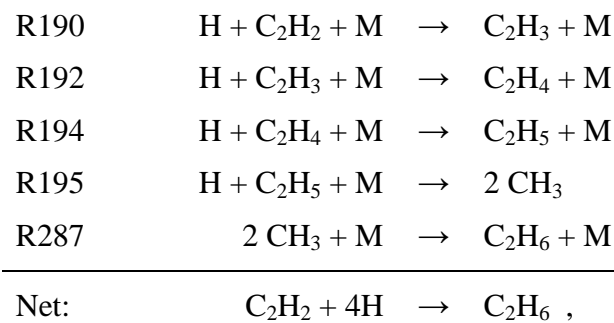

$$
\begin{array}{lrcl}
R190 & H + C_2H_2 + M & \rightarrow & C_2H_3 + M \\
R192 & H + C_2H_3 + M & \rightarrow & C_2H_4 + M \\
R194 & H + C_2H_4 + M & \rightarrow & C_2H_5 + M \\
R195 & H + C_2H_5 + M & \rightarrow & 2\,CH_3 \\
R287 & 2\,CH_3 + M & \rightarrow & C_2H_6 + M \\
\hline
\text{Net:} & C_2H_2 + 4H & \rightarrow & C_2H_6 \ ,
\end{array}
$$

where the number in the first column refers to the reaction numbers in the Moses et al. (2005) kinetic mechanism. Ethane is recycled more effectively through this mechanism at higher pressures, but there are several "leaks" out of the mechanism and/or bottlenecks to the mechanism that occur on Uranus, preventing ethane recycling from being very effective. The most obvious "leak" concerns the fate of $CH_3$ once it is produced by reaction R195. On Uranus, methyl radicals are much more likely to form methane than ethane in the photochemically active region, for the reasons discussed above, significantly reducing the effectiveness of the ethane recycling mechanism. Another "leak" in the mechanism occurs because R192 is not the only consequence of the reaction of H with $C_2H_3$ – the reaction R191 (H + $C_2H_3$ = $C_2H_2$ + $H_2$) competes effectively with R192 and can short-circuit the recycling process by sending the products back to acetylene rather than ethane.

The relative importance of the two reactions, R191 and R192, is a critical factor in determining effectiveness of ethane recycling; however, kinetic uncertainties and literature inconsistencies unfortunately exist with regard to the relative effectiveness of the two reactions (see Moses et al. 2005 for further discussion). In fact, these uncertainties were one of the main factors in the development of two different kinetic mechanisms in the Moses et al. study. Their "Model C" kinetic reaction list, which more closely follows theoretical predictions rather than experimental derivations for the relative rates of these reactions, leads to higher relative rates of R191 under Uranian stratospheric conditions than the "Model A" list does. As such, we adopt the Model C kinetics in our new model to minimize the $C_2H_6/C_2H_2$ ratio in the final results. Model C provides a more effective short-circuit of the above ethane recycling pathway and a lower overall $C_2H_6/C_2H_2$ ratio that is more consistent with the Spitzer observations, but even "Model A" predicts a $C_2H_6/C_2H_2$ ratio that would be lower on Uranus than on the other giant planets due to the restrictive pressure regime over which the hydrocarbon photochemistry operates. At these pressures, ethane loss more closely balances its production, and conversion of acetylene into ethane is not hugely effective.

The end result is that sluggish vertical mixing (and resulting low-altitude methane homopause) on Uranus leads to hydrocarbon photochemistry being initiated in a different pressure regime on Uranus than the other giant planets, resulting in different dominant hydrocarbon production and loss mechanisms and much smaller total hydrocarbon column production rates. Both the relative abundances and overall column abundances of the hydrocarbons are affected.

**5.6. Carbon dioxide and its implications**. As is discussed in Feuchtgruber et al. (1999) and Burgdorf et al. (2006), the presence of measurable amounts of oxygen compounds like $CO_2$ or $H_2O$ in the stratosphere of Uranus requires an external source of oxygen to the planet. Possible sources include satellite/ring debris, cometary impacts, or micrometeoroid ablation (Feuchtgruber et al. 1997, 1999; Moses et al. 2000b, Bézard et al. 2002, Lellouch et al. 2002). The relative abundance of $H_2O$ in comparison with CO and $CO_2$ can help provide clues to the source, as one might expect much of the oxygen delivered from cometary impacts to be converted efficiently to CO (and through subsequent chemistry, more $CO_2$) in comparison to the other external sources (Bézard et al. 2002, Lellouch et al. 2002, Moses et al. 2004). With our photochemistry/diffusion models, we are able to reproduce the $CO_2$ emission observed by the IRS (**Fig. 10**) with a model that includes a water influx of 3 x $10^5$ $cm^{-2}$ $s^{-1}$, a CO influx of 5 x $10^4$ $cm^{-2}$ $s^{-1}$, and a $CO_2$ influx of 3 x $10^3$ $cm^{-2}$ $s^{-1}$, but we have not explored parameter space sufficiently to fully constrain these values. These influx rates are consistent with the $CO_2$ upper limits and with the upper end of the range for the $H_2O$ column abundance inferred from ISO observations (Feuchtgruber et al. 1999) for our assumed eddy diffusion coefficients and thermal structure (the latter of which controls where the $H_2O$ and $CO_2$ condense and thus affects the total column abundance). Because $CO_2$ can be produced photochemically from $H_2O$ and CO, this solution to the relative influx rates is not unique.

Carbon monoxide may also have an internal source (e.g., Fegley and Prinn 1986), and we checked whether an internal mole fraction of 1 x $10^{-8}$ (e.g., Encrenaz et al. 2004, Cavalié et al. 2008) in combination with a water influx of 3 x $10^5$ $cm^{-2}$ $s^{-1}$ could photochemically produce the observed abundance of $CO_2$, without an extra external source of CO and $CO_2$ being needed (cf. the Saturn case, Moses et al. 2000b, Ollivier et al. 2000, Cavalié et al. 2010). However, this resulting $CO_2$ profile fell short by more than an order of magnitude in reproducing the Spitzer spectra, which suggests that some external source of $CO_2$ and/or CO is required to explain the observations. Recent Herschel observations of carbon monoxide suggest an even smaller CO mole fraction of 7.1-9 x $10^{-9}$ in the stratosphere of Uranus (Cavalié et al. 2014) and an upper limit of 2.1 x $10^{-9}$ for a vertically uniform distribution in the troposphere and stratosphere of Uranus (Teanby and Irwin 2013). This further suggests that an independent source of $CO_2$ is needed to explain the observations (e.g., $CO_2$ ablated or produced during cometary impacts, or $CO_2$ ablated from grains coming originally from the Kuiper belt, as suggested by Poppe and Horanyi, 2012).

An interesting oddity for Uranus that differs from the other giant planets is that many of the possible external sources would deliver the oxygen compounds to a region above their own (and above the methane) homopause levels. Vapor from the satellites or rings would enter from the top of the atmosphere and continuously diffuse downward, and small ice and dust particles from micrometeoroids (including Kuiper-belt dust) or from within the Uranian system would ablate at moderately high altitudes (Moses 2001) that are still well above the methane homopause on Uranus. If the source is introduced at high altitudes and is quasi-continuous, carbon-bearing oxygen species such as CO and $CO_2$ could be photochemically destroyed, leading to a secondary hydrocarbon production region at altitudes above the homopause level. The resulting high-altitude hydrocarbon column abundances are so small that they are unlikely to be detectable with current instrument sensitivities. Depending on the exact physics of the impact process, only the small-cometary-impact source (i.e., comets with diameters in the range of tens of meters or smaller) might be able to introduce the oxygen at or below the homopause level because the energetic plume splashback from large-cometary impacts is expected to transport the oxygen-bearing material back to high altitudes, as was observed with the Shoemaker-Levy 9 impacts on Jupiter (e.g., Lellouch 1996, Zahnle 1996, Lellouch et al. 2002, Palotai et al. 2011, Orton et al. 2011, Pond et al. 2012), and small particles are expected to ablate at high altitudes (e.g., Moses 1991, 2001; Moses et al. 2000b). Therefore, if the external source is continuous and not due to small comets, oxygen-bearing molecules might persist at altitudes above the methane homopause level on Uranus and dominate the photochemistry, aerosol production, and perhaps thermal structure, in that region (see **Fig. 11**). A joint analysis of both Spitzer and Herschel observations of $CO_2$, CO and $H_2O$ can be used to refine the influx rates cited above for this model and thus establish whether exogenic oxygen-bearing species dominate over hydrocarbons high in the atmosphere of Uranus. If this is true, then $H_2O$ condensation could act as an important source of high-altitude hazes.

**5.7. Other minor constituents: $CH_3C_2H$ and $CH_3$.** The photochemistry of methylacetylene ($CH_3C_2H$) at conditions relevant to the stratosphere of Uranus is very uncertain, so we do not attempt to draw any definitive conclusions from the failure of our photochemical model to provide a good fit to the observed $CH_3C_2H$ emission. Our $CH_3C_2H$ profiles are consistent with the observations given the overall rate-coefficient uncertainties that can affect this molecule (e.g., Dobrijevic et al. 2010). However, the required scaling factor of 0.43 for $CH_3C_2H$ compared with derived scaling factors much closer to unity for the other hydrocarbons suggests that our kinetic scheme overestimates the net relative production of $CH_3C_2H$ as compared with the other hydrocarbons. We have not included $CH_3C_2H$ condensation in the model, which would reduce the gas-phase methylacetylene abundance in the lower stratosphere (see **Fig. 11**). However, models that assume that $CH_3C_2H$ would be confined to its saturated value in the lower stratosphere do not improve the fit significantly, so there is likely a fundamental, as-yet-unidentified, photochemical explanation for our overprediction of the $CH_3C_2H$ abundance. The resolution of this problem must await future modeling.

The $CH_3$ abundance in our photochemical model is less than the $3\sigma$ upper-limit on the column abundance of $CH_3$ ($3.3 \times 10^{12}$ molec-cm$^{-2}$) constrained by the Spitzer spectra shown in **Fig. 9**. The factors that influence the $CH_3$ abundance are discussed more thoroughly in Bézard et al. (1998, 1999). Given the low-altitude homopause level on Uranus, the $CH_3$ abundance is expected (and observed) to be significantly smaller than that of the other giant planets.

**5.8 Constraints on the strength of atmospheric mixing.** Our photochemical models best match the observations when eddy diffusion coefficients are in the ~2200-2600 cm$^2$ s$^{-1}$ range. Our results are consistent with the methane homopause and hydrocarbon profiles used to explain the globally averaged ISO observations (Encrenaz et al. 1998), so we find no evidence for substantial changes to the global-average strength of stratospheric mixing over time the time period between the 1996 ISO observations and the 2007 Spitzer observations. At face value, our derived mixing ratio for $C_2H_2$ at high stratospheric altitudes is significantly greater than that derived from the low-latitude Voyager UVS solar occultations as analyzed by Bishop et al. (1990), suggesting that the $C_2H_2$ abundance (and possibly the stratospheric $K_{zz}$ values) on Uranus might have increased with time since the 1986 Voyager encounter. However, the UVS observations are also very sensitive to atmospheric structure; our derived thermal structure differs from that used in the Bishop et al. (1990) analysis (see Fig. 5A of Paper 1), and we note that the $C_2H_2$ mole fractions derived by Herbert et al. (1987) from the same UVS occultation are actually larger than our nominal model, so firm conclusions about temporal variations in $C_2H_2$ will have to await more detailed comparisons with the UVS occultation light curves.

**6. Summary, Conclusions and Future Work.**

We have analyzed molecular emission features that appear at ~11-17 μm in the global-average Uranus spectra acquired with Spitzer/IRS near the time of the December 2007 equinox. Paper 1 (Orton et al. 2014) more fully discusses the complete 5-37 μm data set and derives the thermal structure from these data. Here we determine the stratospheric molecular abundances needed to reproduce the observed emission features of $CH_4$, $C_2H_2$, $C_2H_6$, $CH_3C_2H$, $C_4H_2$, $CO_2$ and possibly $CH_3$. We use constraints on the temperatures to develop new photochemical models for Uranus to compare with the Spitzer IRS data. The model profiles are scaled, as needed, to provide the best fit to the emission. Some longitudinal variability, which will be analyzed more fully in a subsequent publication, is obvious in the data. These variations complicate the analysis of molecular abundances, as it is currently unclear whether the longitudinal differences result from variations stratospheric temperatures or hydrocarbon abundances.

We find that photochemical models can accurately reproduce both the methane and ethane emission measured in the IRS spectra simultaneously. For our nominal temperature profile, models with a vertically uniform eddy diffusion coefficient of $K_{zz}$ = 2430 +170/-230 cm$^2$ s$^{-1}$ and a tropopause $CH_4$ mole fraction of 1.6 ± 0.3 x 10$^{-5}$ (corresponding to a methane relative humidity of 23 ± 4%), require no scaling of the $C_2H_6$ profile to match the observed IRS spectra. The resulting $C_2H_6$ profile has a peak mole fraction of 1.3 ± 0.1 x 10$^{-7}$ at 0.2 mbar, and a column abundance of 3.1 ± 0.3 x 10$^{16}$ molecule-cm$^{-2}$ above the 10-mbar level. Along with the other hydrocarbons, the emission from acetylene shows apparent longitudinal variability, and we find that the nominal-model $C_2H_2$ profile must be scaled by 0.85±0.07 to fit the longitudinally averaged SL1 data and by 1.05±0.08 to match the longitudinally averaged LL2 data, where averages were made over different longitude regions. The abundances needed for the SL1 emission correspond to a $C_2H_2$ peak mole fraction of 2.4 ± 0.3 x 10$^{-7}$ at 0.2 mbar, for a total column abundance of 6.2±0.6 x 10$^{16}$ above the 10-mbar level. The fit to the LL2 data requires that these abundances be increased by 24%. (Note that the error limits cited here and below include all sources of uncertainty, including those arising from the uncertainties in the non-LTE function and the absolute calibration of the spectrum, which are not included in the error limits given in **Table 2.**)

Due to uncertainties in the chemical reaction mechanism, the abundance profiles of the other hydrocarbons predicted by the model are more uncertain. The diacetylene profile in our nominal model fits the $C_4H_2$ emission feature well, such that the profile only needs to be scaled by a factor of 1.16±0.16. A key to the good fit for $C_4H_2$ is the adopted vapor-pressure expression at low temperatures in the photochemical model (see Appendix B); further measurements of the $C_4H_2$ vapor pressure at low temperatures would greatly aid analyses of infrared data from both Uranus and Neptune. The nominal photochemical model profile of $CH_3C_2H$ (methylacetylene), on the other hand, overpredicts the emission and must be scaled by 0.43±0.08 to fit the data, suggesting that the relatively uncertain $C_3H_x$ kinetics in our chemical mechanism overestimates the net $CH_3C_2H$ production rate relative to the other hydrocarbons. The potential reasons for the model-data mismatch here deserve further study. Methyl radicals ($CH_3$) are marginally detected, at best, which is consistent with our adopted $K_{zz}$ profile and photochemical mechanism. The derived abundance of $CO_2$ (i.e., column abundance of 1.7±0.2 x 10$^{13}$ molecule-cm$^{-2}$ in our nominal model) is about an order of magnitude lower than the upper-limit column abundance inferred from ISO observations (Feuchtgruber et al. 1999). Our photochemical models suggest that exogenic oxygen species could dominate over hydrocarbons at high altitudes on Uranus, and that water condensation could provide an important source of high-altitude hazes. A $CO_2$ influx of $3 \times 10^3$ $cm^{-2}$ $s^{-1}$ can reproduce the $CO_2$ emission feature detected by Spitzer. Table 1 and Appendix A provide further information about the molecular abundances determined from our best-fit model.

Although our observations provide no obvious evidence for changes in the hydrocarbon abundances with time from 1996 to 2007, our derived $C_2H_2$ abundance is much greater than that determined from the 1986 Voyager UVS solar occultation (Bishop et al. 1990, see **Fig. 11**). Further work should be done to determine whether this result represents a true increase with time (which may imply changes in circulation/atmospheric mixing) or whether the $C_2H_2$ profiles and associated thermal structure derived from these IRS observations are consistent with the UVS occultation light curves (as opposed simply to comparing derived abundances). The addition of seasonal forcing to 1D or 2D photochemical models would also be useful in determining whether changes in hydrocarbon abundances with time are expected on purely photochemical grounds, or whether seasonal changes in eddy diffusion coefficients or atmospheric circulation are needed to influence the hydrocarbon column abundances.

Spectroscopic observations from the Herschel Space Telescope (Hartogh et al. 2000) will provide important constraints on the global-average water abundance in the upper atmosphere of Uranus, and observations with the James Webb Space Telescope (Gardner et al. 2006) will provide spatially resolved information on species abundances.

**Acknowledgements**.

We thank NASA's Spitzer Space Telescope program for initial support of the data acquisition, reduction and its initial analysis, and we thank Tom Soifer for Director's Discretionary Time on Spitzer (program #467). This work is based on observations made with the Spitzer Space Telescope, which is operated by the Jet Propulsion Laboratory, California Institute of Technology under a contract with NASA. Support for this work from the Spitzer program was provided by NASA through an award issued by JPL/Caltech. Another portion of our support was provided to JPL/Caltech, from NASA's Planetary Atmospheres program. J. Moses acknowledges support from NASA Planetary Atmospheres grant NNX13AH81G, as well as older grants from the NASA Planetary Atmospheres program. L. Fletcher acknowledges the Oak Ridge Association of Universities for its support during his tenure at the Jet Propulsion Laboratory in the NASA Postdoctoral Program (NPP), together with the Glasstone and Royal Society Research Fellowships during his current tenure at the University of Oxford. F. J. Martin-Torres was supported by the Spanish Economy and Competitivity Ministry (AYA2011-25720 and AYA2012-38707). During his contribution to this work, M. Line was supported by NASA's Undergraduate Student Research Program.

The radiative-transfer calculations were primarily performed on JPL supercomputer facilities, which were provided by funding from the JPL Office of the Chief Information Officer.

We thank Linda Brown, Helmut Feuchtgruber, Tristan Guillot, Mark Hofstadter, Kathy Rages, Larry Sromovsky, Larry Trafton, and J. Cernicharo for helpful and illuminating conversations, J. Schaefer for help in implementing dimer contributions into the $H_2$ collision-induced opacity calculations, as well as Emmanuel Lellouch and an anonymous reviewer for insightful comments and suggestions.

# Tables

**Table 1**. Derived column abundance of stratospheric species in the pressure region from 10 mbar to 0.1 μbar for the nominal model. Uncertainties cited represent not only the fitting uncertainty (including the propagation of uncertainties associated with fitting adjacent spectral features for $CH_3H_2H$ and $C_4H_2$), but also the uncertainty derived from the range of acceptable $f_{CH4}$ - $K_{zz}$ solutions, which are added in quadrature. They also include propagation of systematic uncertainties associated with the non-LTE function and with absolute calibration, which are treated as potential offsets and thus added linearly. Errors arising from uncertainties in the temperature profile are addressed in the text.

| Molecule | Column abundance (molecule-cm$^{-2}$) | Pressure of maximum contribution function (bars) |
|---|---|---|
| $CH_4$ | 4.5 +1.1/-0.8 x $10^{19}$ | 1.8 x $10^{-4}$ |
| $C_2H_2$ | [a]6.2±1.0 x $10^{16}$ | 1.8 x $10^{-4}$ |
| $C_2H_6$ | 3.1±0.4 x $10^{16}$ | 2.1 x $10^{-4}$ |
| $CH_3C_2H$ | 8.6±2.6 x $10^{13}$ | 4.4 x $10^{-4}$ |
| $C_4H_2$ | 1.8±0.3 x $10^{13}$ | 3.7 x $10^{-4}$ |
| $CO_2$ | 1.7±0.4 x $10^{13}$ | 1.4 x $10^{-4}$ |
| $CH_3$ | [b]<3.3 x $10^{12}$ | 1.5 x $10^{-4}$ |

[a]from fit to the SL1 spectrum; a 24% increase is required to fit the LL2 spectrum
[b]3σ upper limit

**Table 2.** Comparison of properties, including abundance distributions, between the nominal model with a uniform vertical eddy diffusion coefficient $K_{zz}$, the nominal model with a uniform positive vertical $K_{zz}$ slope, and the alternative model with a uniform vertical $K_{zz}$. Unlike Table 1, the uncertainties cited in the column abundances (given as 'col-abund' in units of molecule-cm$^{-2}$) and maximum molar fraction (given by $f_{max}$) do <u>not</u> consider uncertainties from absolute calibration or from the non-LTE function, which affect all models equally. For CH$_4$, $f_{max} = f_{CH4}$, the methane mole fraction at the 100-mbar tropopause. The pressure at which the tropospheric CH$_4$ mole fraction equals 3.2%, as a measure of its vertical slope (see **Fig. 5a**), is given by P$_{CH4\_deep}$. Uncertainties in $K_{zz}$, $f_{CH4}$, and the abundance of C$_2$H$_6$ are completely correlated with one another. The pressure of the maximum stratospheric molar fraction is denoted by p$_{max}$. The best-fit scaling of model abundance profile to match the IRS data is given by $f_{scaling}$, whose definition is identically 1.0 for CH$_4$ and C$_2$H$_6$.

| Model: Property | nominal T(p) uniform $K_{zz}$ | nominal T(p) sloped $K_{zz}$ | alternative T(p) uniform $K_{zz}$ |
|---|---|---|---|
| $f_{CH4}$ | 1.6+0.2/-0.1 x 10$^{-5}$ | 2.5+0.3/-0.2 x 10$^{-5}$ | 7.0 ± 1.0 x 10$^{-5}$ |
| $K_{zz}$ (cm$^2$-s$^{-1}$) | 2430+100/-190 | (see **Fig. 11**) | 590 ± 84 |
| CH$_4$: col-abund | 4.8+ 0.6/- 0.3 x 10$^{20}$ | 7.4+ 0.9/- 0.5 x 10$^{20}$ | 2.1± 0.3 x 10$^{21}$ |
| P$_{CH4\_deep}$ (bars) | 1.78 ± 0.20 | 1.83 ± 0.20 | 1.76 ± 0.20 |
| CH$_3$D/CH$_4$ | 3.0 ± 0.2 x 10$^{-4}$ | 3.0 ± 0.2 x 10$^{-4}$ | 2.7 ± 0.2 x 10$^{-4}$ |
| C$_2$H$_2$: $f_{scaling}$ | [a]0.85 ± 0.07 | [a]0.68 ± 0.06 | [a]0.81 ± 0.07 |
| col-abund | [a]6.2 ± 0.5 x 10$^{16}$ | [a]7.1 ± 0.6 x 10$^{16}$ | [a]7.6 ± 0.7 x 10$^{16}$ |
| $f_{max}$ | 2.4 ± 0.2 x 10$^{-7}$ | 2.0 ± 0.2 x 10$^{-7}$ | 2.2 ± 0.2 x 10$^{-7}$ |
| p$_{max}$ | 2.0 x 10$^{-4}$ | 2.0 x 10$^{-4}$ | 3.2 x 10$^{-4}$ |
| C$_2$H$_6$: col-abund | 3.1 ± 0.2 x 10$^{16}$ | 4.5 ± 0.3 x 10$^{16}$ | 4.8 ± 0.7 x 10$^{16}$ |
| $f_{max}$ | 1.3 ± 0.1 x 10$^{-7}$ | 1.2 ± 0.1 x 10$^{-7}$ | 1.2 ± 0.2 x 10$^{-7}$ |
| p$_{max}$ | 2.0 x 10$^{-4}$ | 2.0 x 10$^{-4}$ | 3.2 x 10$^{-4}$ |
| CH$_3$C$_2$H: $f_{scaling}$ | 0.43 ± 0.08 | 0.30 ± 0.06 | 0.40 ± 0.08 |
| col-abund | 8.6 ± 1.6 x 10$^{13}$ | 9.0 ± 1.7 x 10$^{13}$ | 9.7 ± 1.9 x 10$^{13}$ |
| $f_{max}$ | 4.1 ± 0.8 x 10$^{-10}$ | 4.0 ± 0.8 x 10$^{-10}$ | 4.7 ± 0.9 x 10$^{-10}$ |
| p$_{max}$ | 3.2 x 10$^{-4}$ | 4.0 x 10$^{-4}$ | 5.0 x 10$^{-4}$ |
| C$_4$H$_2$: $f_{scaling}$ | 1.16 ± 0.14 | 0.74 ± 0.09 | 0.94 ± 0.11 |
| col-abund | 1.8 ± 0.2 x 10$^{13}$ | 2.1 ± 0.2 x 10$^{13}$ | 2.1 ± 0.2 x 10$^{13}$ |
| $f_{max}$ | 1.3 ± 0.2 x 10$^{-10}$ | 1.4 ± 0.3 x 10$^{-10}$ | 1.5 ± 0.2 x 10$^{-10}$ |
| p$_{max}$ | 4.0 x 10$^{-4}$ | 5.0 x 10$^{-4}$ | 6.3 x 10$^{-4}$ |
| CO$_2$: col-abund | 1.7 ± 0.2 x 10$^{13}$ | 2.0 ± 0.2 x 10$^{13}$ | 2.1 ± 0.2 x 10$^{13}$ |
| $f_{max}$ | 9.0 ± 1.1 x 10$^{-11}$ | 8.6 ± 1.1 x 10$^{-11}$ | 8.2 ± 1.1 x 10$^{-11}$ |
| p$_{max}$ | 6.3 x 10$^{-5}$ | 5.0 x 10$^{-5}$ | 7.9 x 10$^{-7}$ |
| influx rate (cm$^{-2}$ s$^{-1}$) | 3.0 ± 0.3 x 10$^{3}$ | 2.7 ± 0.3 x 10$^{3}$ | 3.5 ± 0.4 x 10$^{3}$ |
| CH$_3$: $f_{scaling}$ | [b]<2.1 | [b]<2.1 | [b]<4.0 |
| col-abund | [b]<3.3 x 10$^{12}$ | [b]<3.3 x 10$^{12}$ | [b]<4.6 x 10$^{12}$ |
| $f_{max}$ | [b]<7.1 x 10$^{-11}$ | [b]<6.8 x 10$^{-11}$ | [b]<8.1 x 10$^{-11}$ |
| p$_{max}$ | 1.3 x 10$^{-4}$ | 1.3 x 10$^{-4}$ | 2.5 x 10$^{-4}$ |

[a]from fit to the SL1 spectrum; a 24% increase is required to fit the LL2 spectrum
[b]3σ upper limit; the plot in **Fig. 11** plots the best-fit profile for $f_{scaling}$=1.5

# Figures

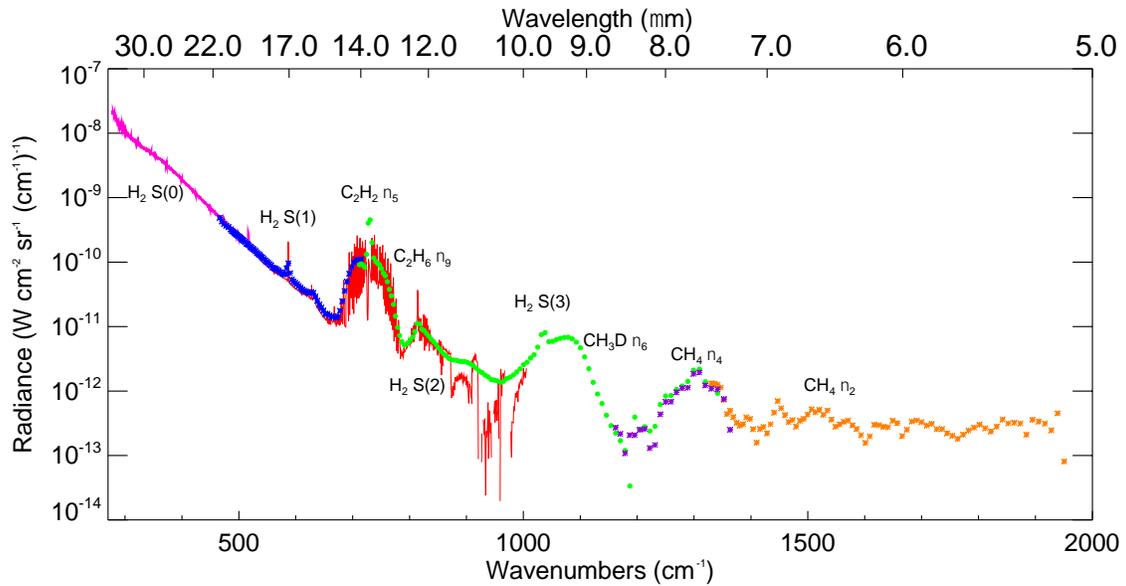

**Figure 1**. The spectrum of Uranus, based on the mean of samples at all longitudes, as observed by all modules the Spitzer IRS on 2007 December 16-17. Different modules are coded in different colors and shapes: SL2 - orange asterisks, SL "bonus order" – purple asterisks, SL1 - green points, LL2 – blue asterisks, SH – red lines, LH – purple points. Radiance is shown *vs.* wavenumber with wavelength given as the upper abscissa. Some spectral features are identified for clarity. We use SL1, LL2 and SH spectra in this study.

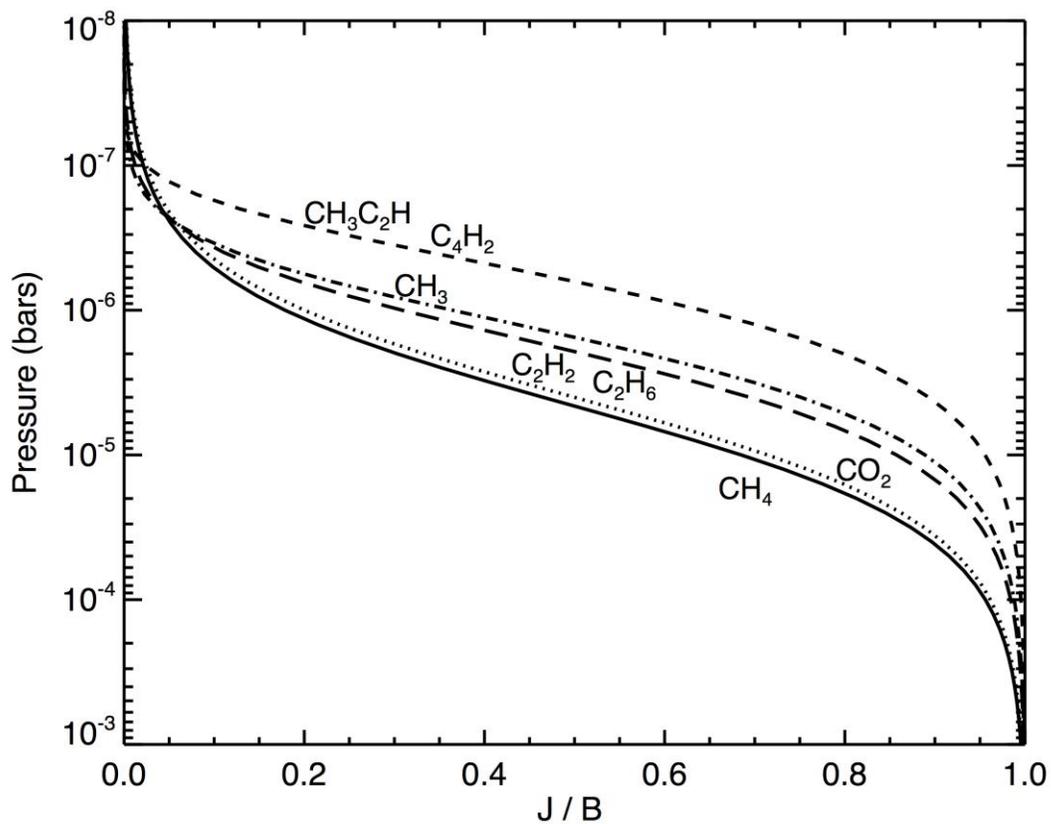

**Figure 2**. Ratio of the source function to the Planck function, J/B, for the molecules under study here. This ratio for $CH_4$ is shown by the solid line and for $CO_2$ by the dotted line. The ratios for $C_2H_2$ and $C_2H_6$ are indistinguishable and are shown by the long dashed line. Similarly the ratios for $CH_3C_2H$ and $C_4H_2$ are indistinguishable and are shown by the short dashed line. The ratio for $CH_3$ is shown by the alternating dot-dashed line.

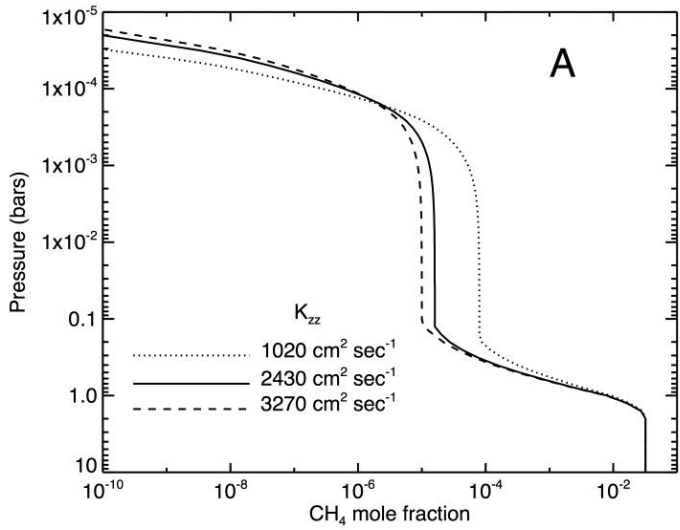

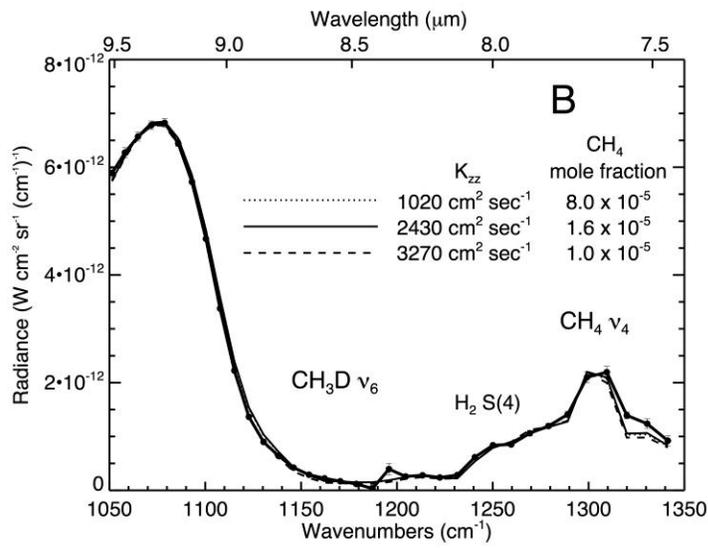

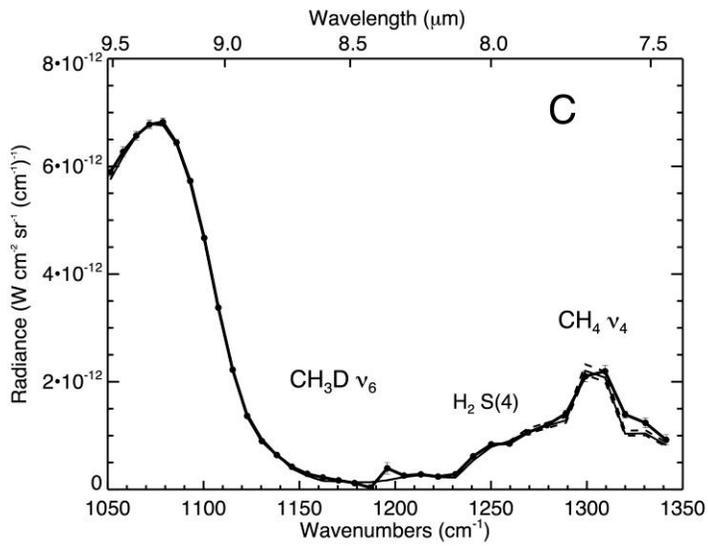

**Figure 3**. (Preceding page) *Panel A*. Best-fit profiles for the vertical distribution of methane and their dependence on the eddy diffusion coefficient. The abundance is indicated at each level by the mole fraction. The value at the temperature minimum remains roughly constant throughout the overlying warmer stratosphere, with an eventual sharper falloff at lower pressures due to molecular diffusion. The model with a tropopause $CH_4$ mole fraction $f_{CH4} = 8 \times 10^{-5}$ is used only to illustrate equally valid fits to the $CH_4$ emission spectrum; it corresponds to ~114% relative humidity. *Panel B*. Comparison of the spectra corresponding to the models shown in Panel A with the 7.4 – 9.5 μm SL1 spectrum. SL1 data are plotted with filled black circles and the dark solid lines, with error bars (which are generally smaller than the diameter of the dark circles). *Panel C*. Sensitivity of the spectral models to the tropopause $CH_4$ mole fraction. The dashed lines show ±1σ perturbations of the $CH_4$ tropopause mole fraction $f_{CH4}$ (±0.2 x 10$^{-5}$) around the model denoted by the solid line with $f_{CH4} = 1.6 \times 10^{-5}$ (~23% relative humidity) for $K_{zz} = 2430$ cm$^2$ sec$^{-1}$ shown in panel A. Because $f_{CH4}$ and $K_{zz}$ they are highly correlated, the dashed lines also illustrate the equivalent ±1σ perturbations of $K_{zz}$ (±100 cm$^2$ s$^{-1}$) around its best-fit value for $f_{CH4} = 1.6 \times 10^{-5}$. Note that the uncertainty values cited here represent only the fit at a single point; a full inventory of the fitting uncertainty must include the range of acceptable $f_{CH4}$ and $K_{zz}$ values, as illustrated in Figs. 5 and 6. Data are illustrated as in Panel B.

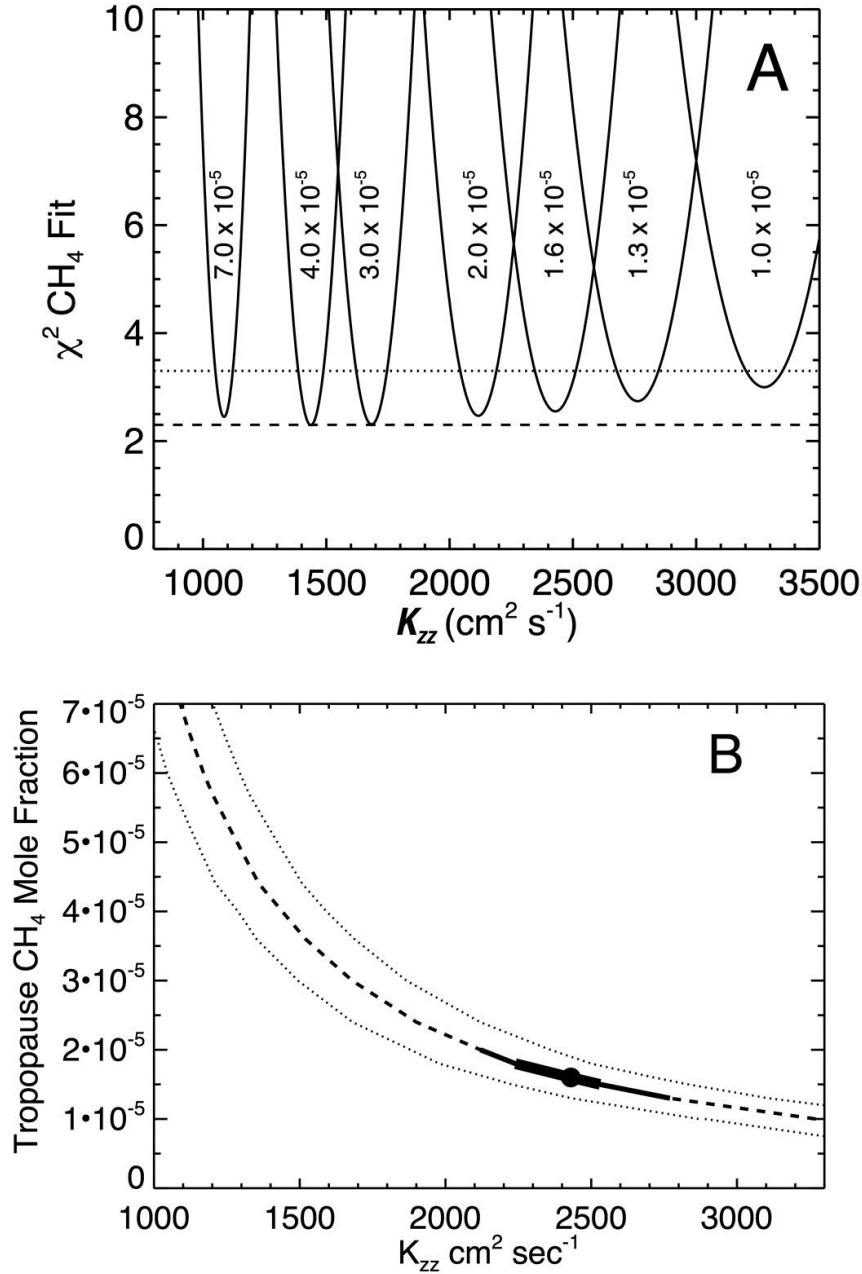

**Figure 4**. *Panel A*. Goodness of fit ($\chi^2$) for a family of solutions for a range of $CH_4$ mole fractions as a function of eddy diffusion coefficients $K_{zz}$. Note that none of the minima for each of the $CH_4$ mole fraction cases is higher than the minimum $\chi^2 + 1$, thus this fit cannot differentiate between these models. *Panel B*. Best-fit values for $K_{zz}$ vs. $CH_4$ tropospheric mole fraction. The suite of best fits shown in Panel A is given by the dashed line and ±1σ fitting uncertainties on either side of it by the dotted lines. When combined with $C_2H_6$ constraints shown in Fig. 6, this suite is narrowed to the range given by the thin solid line, which defines the ±3-σ boundaries of the fit, and the thick solid line, which defines the ±1σ boundaries of the fit. The filled circle shows the best fit. Note that Figures 3A and 3B illustrate the end members of this family, as well as the best fit.

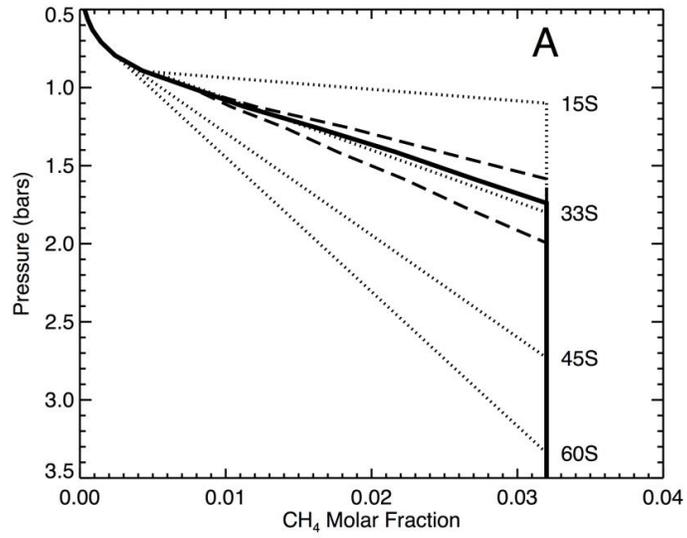
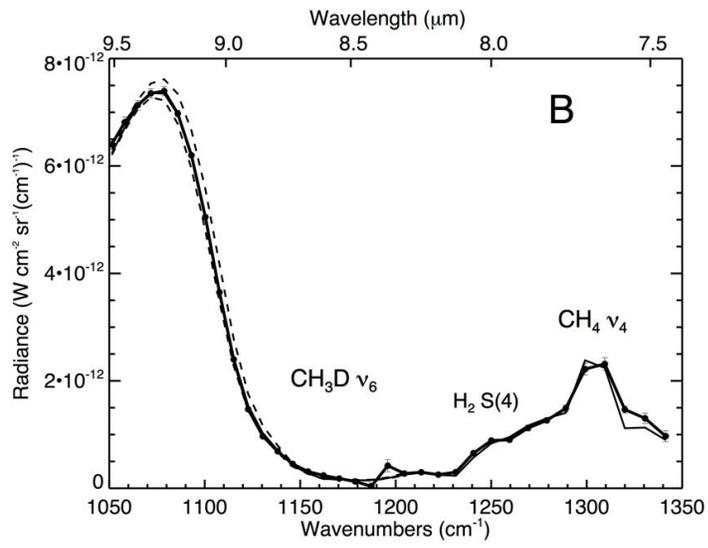
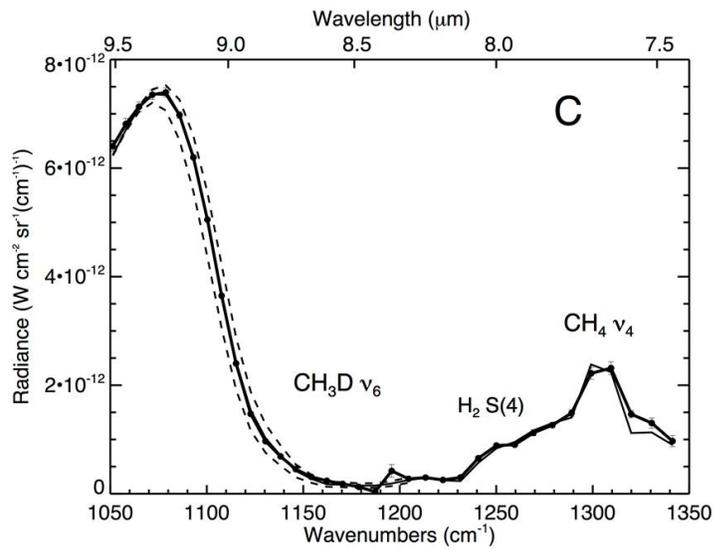

**Figure 5**. (Preceding page) *Panel A*. The best-fit profile for the vertical distribution of methane in the troposphere of Uranus (solid line). The dashed lines illustrate the ±2σ limits on that distribution. The models for $CH_4$ distribution in Uranus derived from HST STIS illustrated by Karkoschka and Tomasko (2011) are also shown schematically (dotted lines). We note that our global average is close to their distribution at 33°S. *Panel B*. The spectra corresponding to the best-fit model and its uncertainties. The same linestyles are used as in Panel A. *Panel C*. The model spectrum corresponding to the best fit of the $CH_3D/CH_4$ ratio, $3.0+0.2/-0.1 \times 10^{-4}$. The dotted lines show its ±7σ uncertainties for clarity. Note that none of the differences between these model cases influences the 7.4-8.3 μm $CH_4$ emission spectrum. Data in Panels B and C are plotted just as in Figure 3.

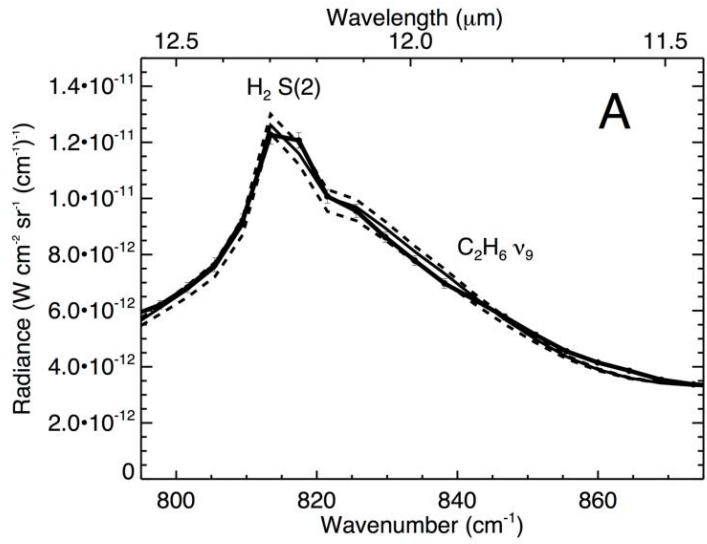
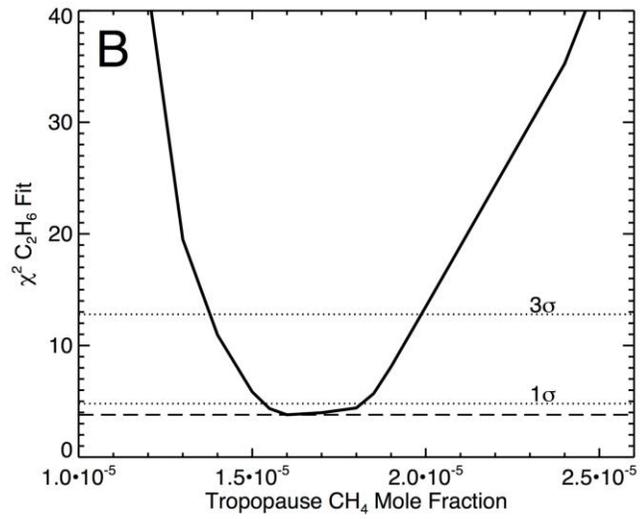
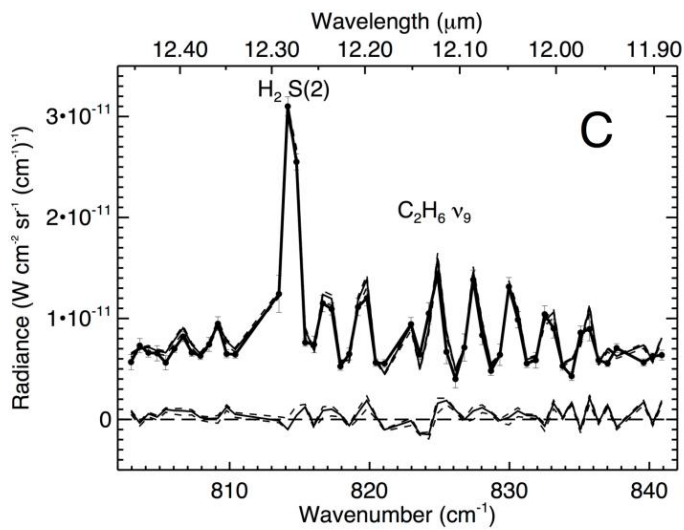

**Figure 6**. (Preceding page). *Panel A*. Fit to the $C_2H_6$ feature in the SL1 spectrum for the best-fit nominal model and for the models defining the $\pm 3\sigma$ boundaries of the fit. *Panel B*. Goodness of fit ($\chi^2$) as a function of the $CH_4$ tropopause mole fraction. Its values are completely correlated with the corresponding value of $K_{zz}$ shown in Panel B of **Fig. 5**. The minimum $\chi^2 + 1$ and minimum $\chi^2 + 9$ levels are shown, from which the $\pm 1\sigma$ and $\pm 3\sigma$ boundaries shown in Figure 5C are derived (Bevington et al. 1992). *Panel C*. The same $C_2H_6$ feature in the SH spectrum, together with the corresponding spectrum of the best-fit model (derived from the SL1 spectrum shown in Panel B) together with spectra for models corresponding to $\pm 1\sigma$ fits (dashed lines). Residuals to the fit, shown in the same panel around zero radiance using the same linestyles, do not indicate any additional spectral component. The central portion of this figure, illustrating the $H_2$ S(2) feature, is shown in Figure 13, panels E and F, of Paper 1.

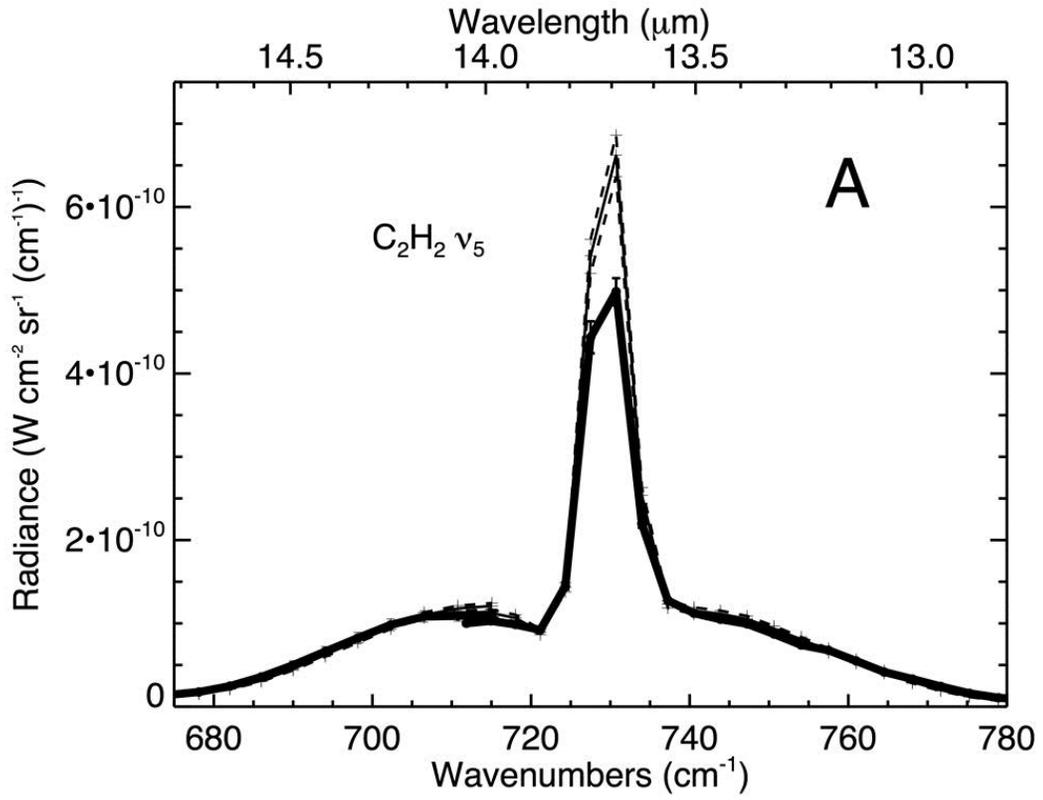

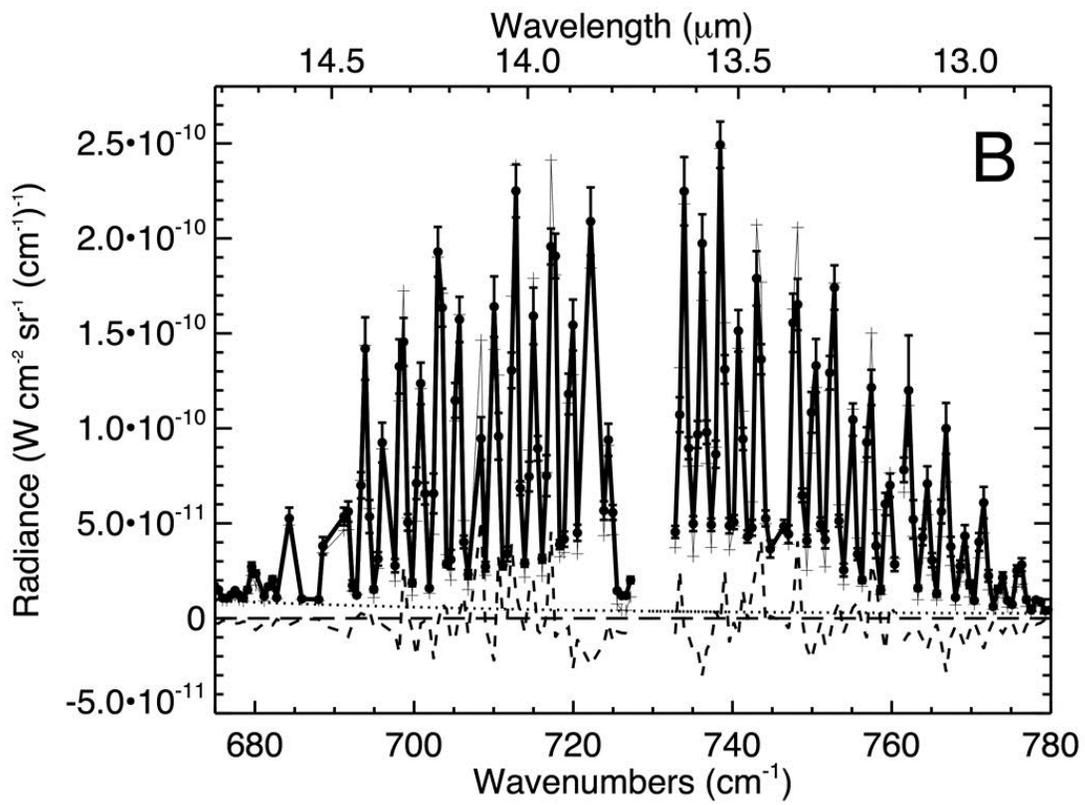

**Figure 7**. (Preceding page) Fitting the $C_2H_2$ abundance. *Panel A*. Comparison of model spectra for the $C_2H_2$ stratospheric abundance with the photometric SL1 and LL2 spectra. The thick solid line and filled circles plot the LL2 spectrum at wavelengths equal to or longer than 14.00 μm (715 cm$^{-1}$), and the thick solid line and filled circles plot the SL1 spectrum at wavelengths shorter than 14.04 μm (712 cm$^{-1}$). The thinner solid lines show the best fit to the LL2 and SL1 spectra for the nominal model ($K_{zz}$ = 2430 cm$^2$ sec$^{-1}$ and $CH_4$ tropopause mole fraction of 1.6 x 10$^{-5}$). The predicted $C_2H_2$ abundance profile from the nominal photochemical model must be scaled by 0.85±0.07 to fit the SL1 data and by 1.05±0.08 to match the LL2 data. Model spectra representing these 1σ departures from the best-fit scaling factors are shown by dotted lines for the LL2 spectrum and dashed lines for the SL1 spectrum. *Panel B*. Best fits to the portion of the SH spectrum showing $C_2H_2$ features in orders 14, 15, and 16. The data are shown by the thick solid lines with filled circles. The best-fit nominal model is shown by the solid lines, whose residuals from the observed spectrum are shown by the dashed lines around zero radiance (noted by the long-dashed line). The $H_2$ collision-induced component of the spectrum is shown by the dotted curve. The scaling factors were the same as for the LL2 and SL1 spectra shown in Panel A.

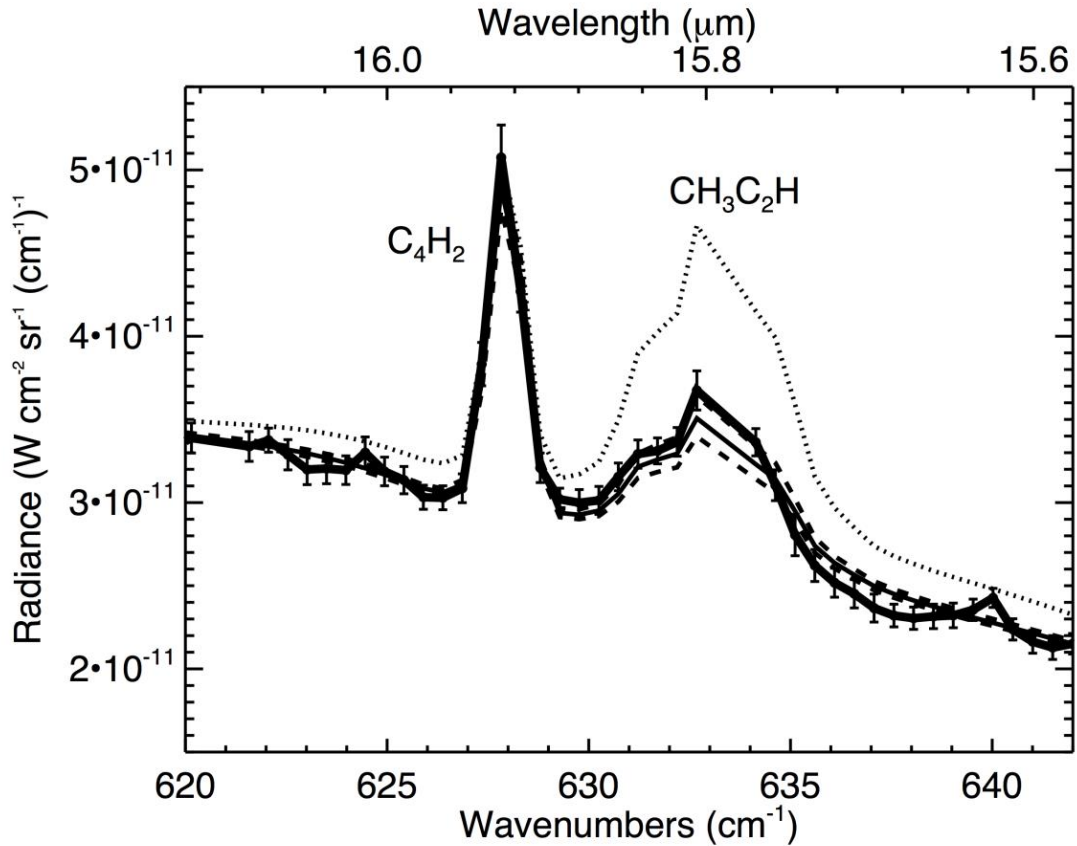

**Figure 8**. Comparison of models for the stratospheric abundances of diacetylene ($C_4H_2$) and methylacetylene ($CH_3C_2H$) with the observed SH spectrum in order 13. The abundance profiles were derived from scaling the abundance profiles in the nominal model. The SH spectrum is calibrated to fit the lower-resolution LL2 spectrum. The thick solid line with filled circles represents the observed spectrum. The regular solid line represents the best-fit spectrum, which is surrounded above and below by dashed lines that represent the spectra corresponding to scaling factors that are 1σ from the best fit. The dotted line represents the spectrum corresponding to the spectral feature of $CH_3C_2H$ if its abundance were not scaled by a best-fit factor of 0.43. The spectral feature near 15.63 μm (640 cm$^{-1}$), which appears in the spectrum at each longitude, is

unidentified

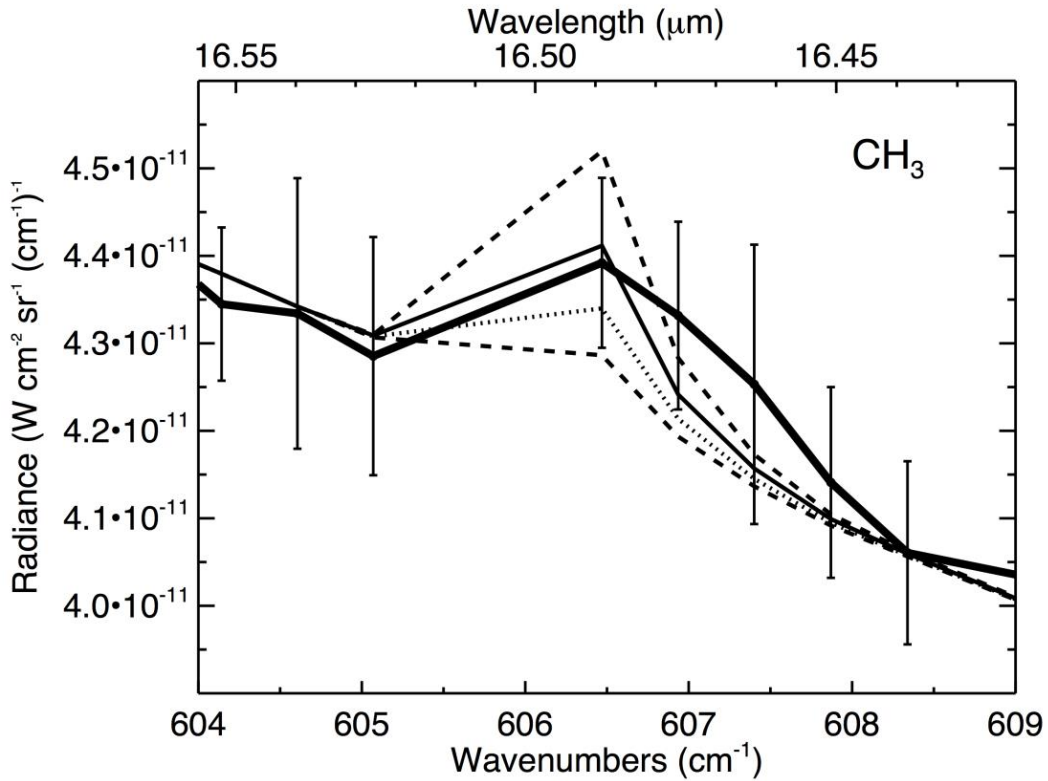

**Figure 9**. Comparison of spectral models with a possible methyl ($CH_3$) $\nu_2$ fundamental band emission feature detected in SH order 12. The observed spectrum is represented by filled black circles connected by the thick solid black line. The regular solid line represents the best-fit model spectrum, with model spectra representing 1σ scaling factors identified above and below the nominal model by dashed lines. The dotted line represents the spectrum if the abundance of $CH_3$ were not scaled by a best-fit factor of 1.50. We note that the fit to the spectrum does not improve by adopting a higher spectral resolution. The uncertainty of the fit is nearly half of the value itself; with such a low signal-to-noise ratio, we cite only a 3σ upper limit for the scaling factor of 2.1.

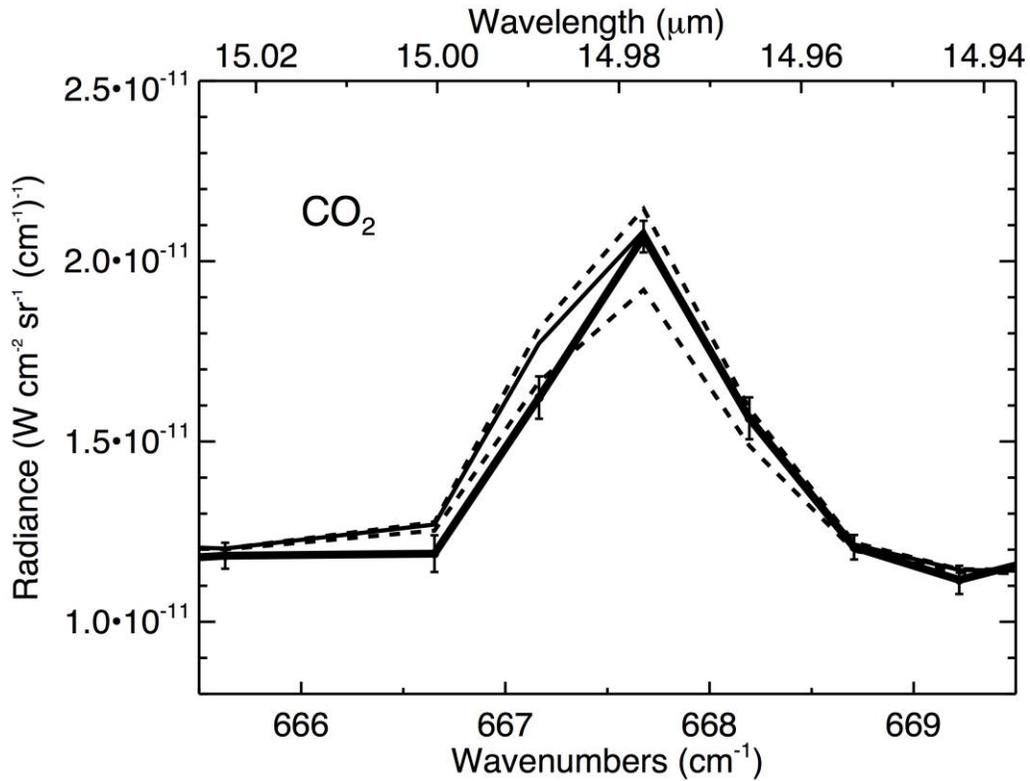

**Figure 10**. Model fits to the $\nu_2$ fundamental of $CO_2$, using an unscaled photochemical model vertical profile of $CO_2$ in which the carbon dioxide was introduced to the atmosphere through the ablation of icy Kuiper-belt dust grains (with a pressure dependence described by Moses, 2001). The thick solid line with filled circles represents the SH observations in order 14. Dashed lines above and below this spectrum represent $1\sigma$ differences of $\pm 0.08$ from the best-fit mole-fraction profile, shown in **Fig.** 12.

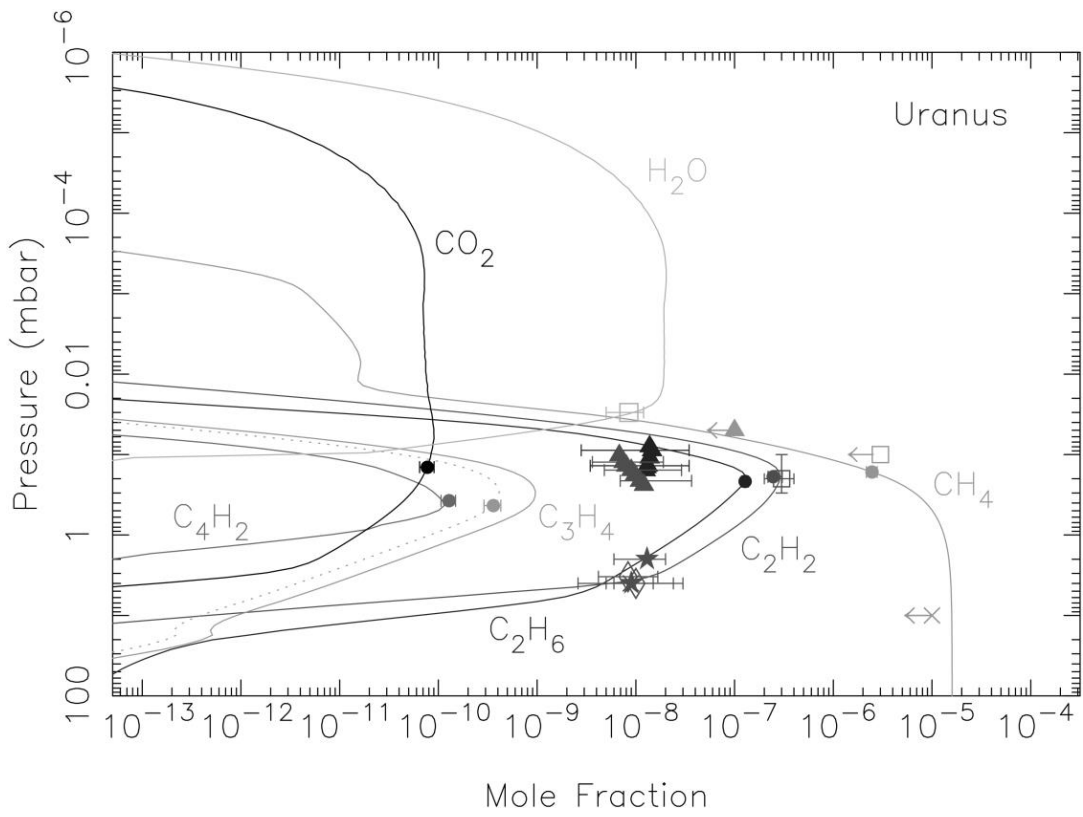
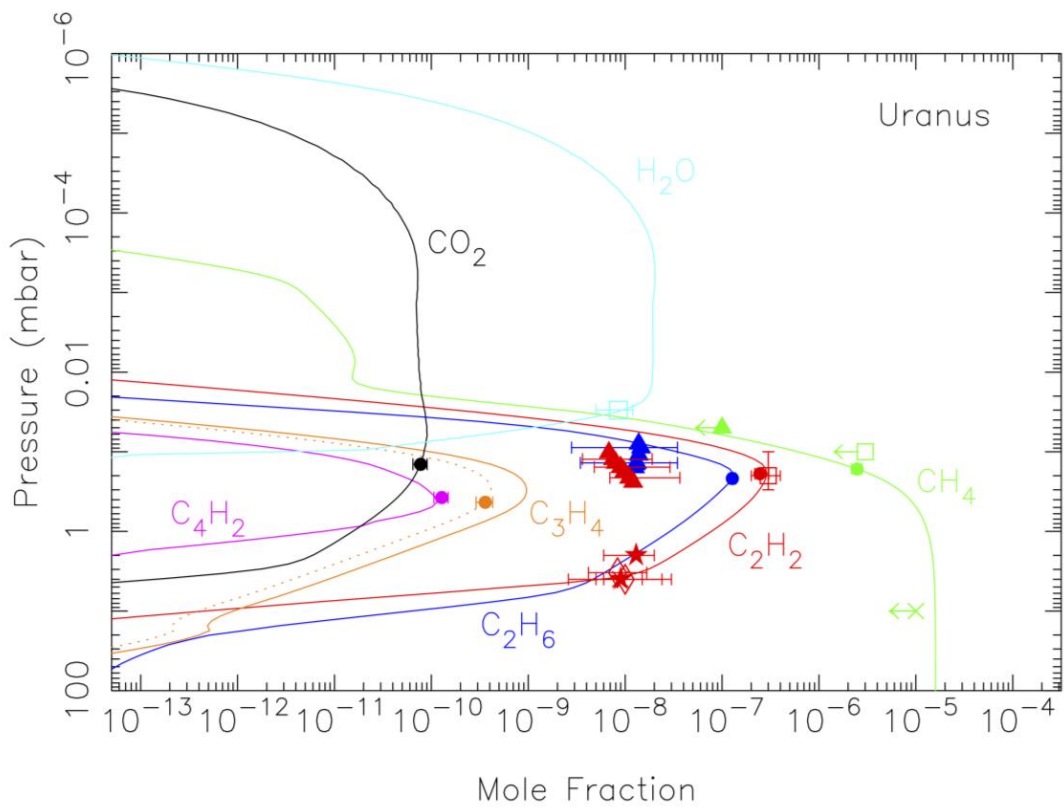

**Figure 11**. (preceding page: color version is for the on-line article) Derived global-mean mole fractions for $CH_4$, $C_2H_2$, $C_2H_6$, $CH_3C_2H$, $C_4H_2$, and $CO_2$ plotted as solid circles with associated error bars from our analysis of the Spitzer IRS spectra. Observational data from other sources are included in the plot: open squares from ISO data (Encrenaz et al. 1998, Feuchtgruber et al. 1999); solid squares from ground-based infrared data (Orton et al. 1987); solid triangles from Voyager UVS occultation data (Bishop et al. 1990, with the $CH_4$ upper limit from Herbert et al. 1987); solid stars from Voyager UVS solar reflection data (Yelle et al. 1989); open diamonds from IUE observations (Encrenaz et al. 1986, Caldwell et al. 1988). The solid lines represent the unscaled mole fraction profiles from our nominal photochemical model that assumes $K_{zz} = 2430$ cm$^2$ s$^{-1}$ and a $CH_4$ mole fraction 1.6 x 10$^{-5}$ at the tropopause. Note from a comparison of these model profiles (solid lines) with the Spitzer data (solid circles) that the profiles of $C_2H_2$, $C_2H_6$, $C_4H_2$, and $CO_2$ need some minor scaling to provide a best fit to our Spitzer spectra. Note that the solid circles representing Spitzer data are plotted at pressures that correspond to the peaks of the respective contributions functions. The $CH_3C_2H$ profile needed more significant scaling, and the resulting scaled profile is shown by a dotted line. Adopting the alternative "isothermal" profile would increase these abundances by factors of several, with the best-fit $CH_4$ tropopause abundance corresponding to 100% saturation. The column abundances corresponding to the best-fit scaled profiles are given in **Table 1** and other details are summarized in **Table 2**, with full tables of all temperature and abundance profiles in **Appendix A**.

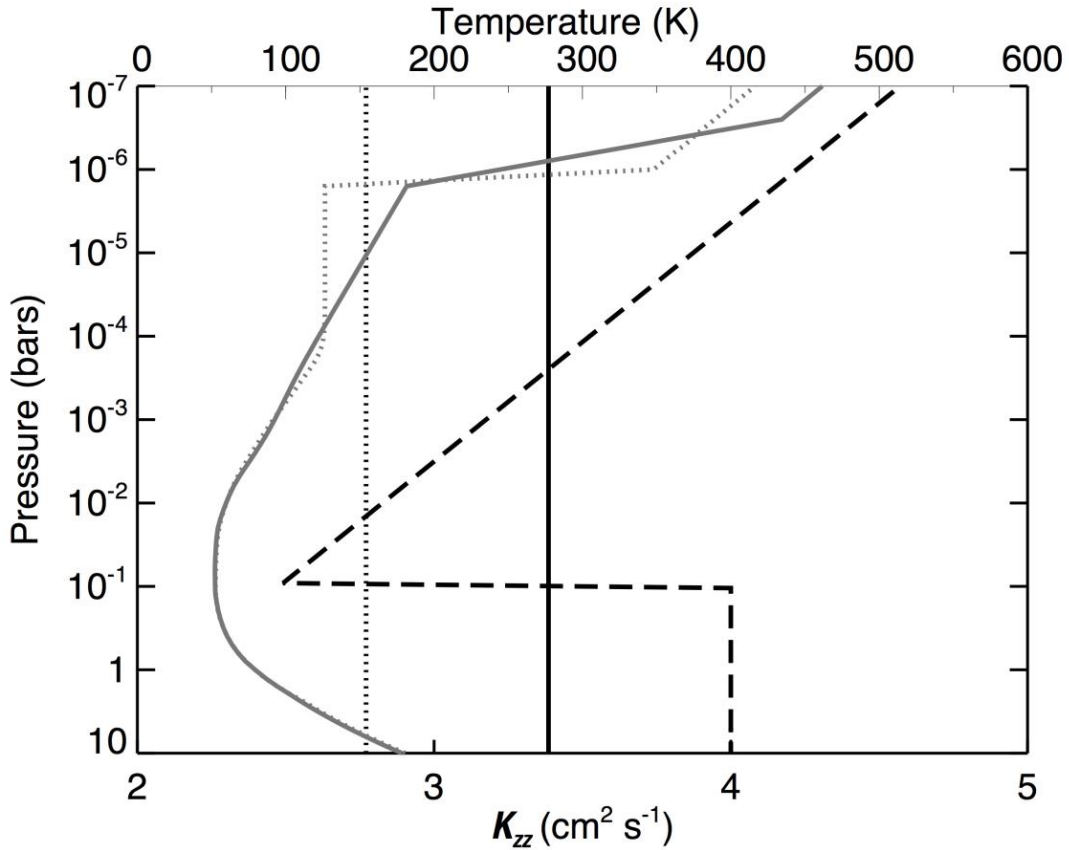

**Figure 12**. Vertical profiles of the eddy diffusion coefficient $K_{zz}$, co-plotted with corresponding temperature profiles. The uniform and sloped profiles of $K_{zz}$ associated with the best spectral fits to the nominal temperature profile are shown by the thick black line and the dashed line, respectively. The uniform $K_{zz}$ associated with the alternative temperature profile is shown by the black dotted line. The thick gray line shows the temperature profile of the nominal model, and the dotted gray line shows the temperature profile of the alternative model. The $K_{zz}$ values in the upper troposphere and the base of the stratosphere are loosely based on results for Jupiter and the other outer planets (Bézard et al. 2002, Moses et al 2005), but they are not well constrained.

**Appendix A.** Table of pressure vs. (col. 2) temperature and (cols. 3-9) best-fit volume mixing ratios for the nominal model for pressures equal to or greater than 1 x 10$^{-7}$ bars. The He/H$_2$ ratio is assumed to be 15/85 by number (Conrath et al. 1987). A digital version of this table is available in the on-line Supplemental Material.

| P(bar) | T (K) | mole fraction | | | | | | |
|---|---|---|---|---|---|---|---|---|
| | | CH$_4$ | C$_2$H$_2$ | C$_2$H$_6$ | CH$_3$C$_2$H | C$_4$H$_2$ | CH$_3$ | CO$_2$ |
| 1.00e+01 | 178.51 | 3.20e-02 | 1.13e-11 | 4.86e-11 | 2.18e-16 | 1.86e-15 | 9.62e-33 | 9.94e-15 |
| 8.91e+00 | 172.05 | 3.20e-02 | 1.22e-11 | 5.24e-11 | 2.36e-16 | 2.07e-15 | 1.82e-30 | 1.07e-14 |
| 7.94e+00 | 165.77 | 3.20e-02 | 1.32e-11 | 5.66e-11 | 2.56e-16 | 2.30e-15 | 3.44e-28 | 1.16e-14 |
| 7.08e+00 | 159.78 | 3.20e-02 | 1.43e-11 | 6.11e-11 | 2.77e-16 | 2.56e-15 | 6.48e-26 | 1.25e-14 |
| 6.31e+00 | 153.94 | 3.20e-02 | 1.54e-11 | 6.60e-11 | 3.00e-16 | 2.84e-15 | 1.22e-23 | 1.35e-14 |
| 5.62e+00 | 148.24 | 3.20e-02 | 1.66e-11 | 7.13e-11 | 3.25e-16 | 3.16e-15 | 2.31e-21 | 1.46e-14 |
| 5.01e+00 | 142.70 | 3.20e-02 | 1.80e-11 | 7.70e-11 | 3.53e-16 | 3.51e-15 | 4.36e-19 | 1.58e-14 |
| 4.47e+00 | 137.30 | 3.20e-02 | 1.95e-11 | 8.28e-11 | 3.97e-16 | 4.89e-15 | 3.22e-18 | 1.70e-14 |
| 3.98e+00 | 132.07 | 3.20e-02 | 2.12e-11 | 8.89e-11 | 4.46e-16 | 6.80e-15 | 2.38e-17 | 1.83e-14 |
| 3.55e+00 | 127.00 | 3.20e-02 | 2.33e-11 | 9.47e-11 | 5.60e-16 | 1.27e-14 | 2.75e-17 | 1.96e-14 |
| 3.16e+00 | 122.08 | 3.20e-02 | 2.57e-11 | 1.01e-10 | 7.02e-16 | 2.36e-14 | 3.18e-17 | 2.10e-14 |
| 2.82e+00 | 117.32 | 3.20e-02 | 2.94e-11 | 1.06e-10 | 9.63e-16 | 5.21e-14 | 2.42e-17 | 2.24e-14 |
| 2.51e+00 | 112.70 | 3.20e-02 | 3.37e-11 | 1.11e-10 | 1.32e-15 | 1.15e-13 | 1.85e-17 | 2.39e-14 |
| 2.24e+00 | 108.23 | 3.20e-02 | 3.94e-11 | 1.15e-10 | 1.44e-15 | 1.93e-13 | 1.68e-17 | 2.54e-14 |
| 2.00e+00 | 103.90 | 3.20e-02 | 4.61e-11 | 1.19e-10 | 1.56e-15 | 3.24e-13 | 1.53e-17 | 2.70e-14 |
| 1.78e+00 | 99.58 | 2.95e-02 | 5.28e-11 | 1.23e-10 | 1.21e-15 | 1.38e-13 | 1.82e-17 | 2.87e-14 |
| 1.58e+00 | 95.26 | 2.72e-02 | 6.04e-11 | 1.27e-10 | 9.31e-16 | 5.91e-14 | 2.16e-17 | 3.04e-14 |
| 1.41e+00 | 91.29 | 2.12e-02 | 6.76e-11 | 1.30e-10 | 6.53e-16 | 9.35e-15 | 2.85e-17 | 3.20e-14 |
| 1.26e+00 | 87.32 | 1.65e-02 | 7.57e-11 | 1.33e-10 | 4.58e-16 | 1.48e-15 | 3.75e-17 | 3.38e-14 |
| 1.12e+00 | 82.85 | 1.15e-02 | 8.30e-11 | 1.37e-10 | 3.29e-16 | 2.06e-16 | 4.55e-17 | 3.34e-14 |
| 1.00e+00 | 80.21 | 7.94e-03 | 9.11e-11 | 1.40e-10 | 2.36e-16 | 2.87e-17 | 5.52e-17 | 3.31e-14 |
| 8.91e-01 | 76.90 | 4.47e-03 | 8.83e-11 | 1.44e-10 | 1.78e-16 | 1.17e-17 | 5.99e-17 | 1.86e-14 |
| 7.94e-01 | 73.60 | 2.52e-03 | 8.56e-11 | 1.47e-10 | 1.34e-16 | 4.74e-18 | 6.49e-17 | 1.04e-14 |
| 7.08e-01 | 71.04 | 1.51e-03 | 4.11e-11 | 1.46e-10 | 1.10e-16 | 1.54e-18 | 4.79e-17 | 2.72e-15 |
| 6.31e-01 | 68.49 | 8.99e-04 | 1.97e-11 | 1.45e-10 | 9.06e-17 | 5.01e-19 | 3.54e-17 | 7.12e-16 |
| 5.62e-01 | 66.39 | 5.54e-04 | 4.55e-12 | 1.07e-10 | 8.41e-17 | 3.45e-20 | 1.76e-17 | 1.60e-16 |
| 5.01e-01 | 64.29 | 3.42e-04 | 1.05e-12 | 7.84e-11 | 7.81e-17 | 2.37e-21 | 8.77e-18 | 3.60e-17 |
| 4.47e-01 | 62.63 | 2.24e-04 | 2.61e-13 | 3.69e-11 | 8.20e-17 | 3.57e-22 | 4.48e-18 | 9.51e-18 |
| 3.98e-01 | 60.96 | 1.47e-04 | 6.49e-14 | 1.74e-11 | 8.61e-17 | 5.37e-23 | 2.29e-18 | 2.51e-18 |
| 3.55e-01 | 59.66 | 1.04e-04 | 2.10e-14 | 7.60e-12 | 9.95e-17 | 6.03e-23 | 1.41e-18 | 8.66e-19 |
| 3.16e-01 | 58.35 | 7.36e-05 | 6.83e-15 | 3.33e-12 | 1.15e-16 | 6.76e-23 | 8.70e-19 | 2.99e-19 |
| 2.82e-01 | 57.29 | 5.57e-05 | 2.69e-15 | 1.61e-12 | 1.08e-16 | 2.93e-22 | 8.40e-19 | 1.24e-19 |
| 2.51e-01 | 56.24 | 4.21e-05 | 1.06e-15 | 7.83e-13 | 1.02e-16 | 1.27e-21 | 8.12e-19 | 5.17e-20 |
| 2.24e-01 | 55.43 | 3.38e-05 | 4.90e-16 | 4.29e-13 | 6.74e-17 | 3.74e-22 | 4.81e-19 | 2.50e-20 |
| 2.00e-01 | 54.62 | 2.71e-05 | 2.27e-16 | 2.36e-13 | 4.45e-17 | 1.10e-22 | 2.85e-19 | 1.21e-20 |
| 1.78e-01 | 54.07 | 2.33e-05 | 1.30e-16 | 1.52e-13 | 3.14e-17 | 4.47e-23 | 1.48e-19 | 7.11e-21 |
| 1.59e-01 | 53.52 | 2.00e-05 | 7.46e-17 | 9.75e-14 | 2.21e-17 | 1.82e-23 | 7.65e-20 | 4.18e-21 |
| 1.41e-01 | 53.10 | 1.79e-05 | 4.89e-17 | 7.08e-14 | 1.81e-17 | 1.06e-23 | 9.26e-20 | 2.86e-21 |
| 1.26e-01 | 52.68 | 1.60e-05 | 3.20e-17 | 5.13e-14 | 1.49e-17 | 6.17e-24 | 1.12e-19 | 1.95e-21 |
| 1.12e-01 | 52.55 | 1.60e-05 | 2.87e-17 | 4.66e-14 | 2.39e-17 | 1.04e-23 | 1.38e-19 | 1.74e-21 |

Appendix A, Continued.

| P(bar) | T (K) | mole fraction | | | | | | |
|---|---|---|---|---|---|---|---|---|
| | | $CH_4$ | $C_2H_2$ | $C_2H_6$ | $CH_3C_2H$ | $C_4H_2$ | $CH_3$ | $CO_2$ |
| 1.00e-01 | 52.42 | 1.60e-05 | 2.57e-17 | 4.23e-14 | 3.84e-17 | 1.74e-23 | 1.71e-19 | 1.56e-21 |
| 8.91e-02 | 52.42 | 1.60e-05 | 2.54e-17 | 4.10e-14 | 7.31e-17 | 4.53e-23 | 2.91e-19 | 1.53e-21 |
| 7.94e-02 | 52.42 | 1.60e-05 | 2.51e-17 | 3.98e-14 | 1.39e-16 | 1.18e-22 | 4.95e-19 | 1.50e-21 |
| 7.08e-02 | 52.38 | 1.60e-05 | 2.66e-17 | 4.14e-14 | 2.84e-16 | 4.01e-22 | 1.22e-18 | 1.58e-21 |
| 6.31e-02 | 52.35 | 1.60e-05 | 2.82e-17 | 4.31e-14 | 5.81e-16 | 1.36e-21 | 3.02e-18 | 1.66e-21 |
| 5.62e-02 | 52.42 | 1.60e-05 | 3.36e-17 | 4.89e-14 | 1.20e-15 | 4.53e-21 | 8.91e-18 | 1.94e-21 |
| 5.01e-02 | 52.49 | 1.60e-05 | 4.01e-17 | 5.54e-14 | 2.46e-15 | 1.51e-20 | 2.63e-17 | 2.26e-21 |
| 4.47e-02 | 52.64 | 1.60e-05 | 5.19e-17 | 6.69e-14 | 4.82e-15 | 4.11e-20 | 6.61e-17 | 2.84e-21 |
| 3.98e-02 | 52.79 | 1.60e-05 | 6.71e-17 | 8.06e-14 | 9.46e-15 | 1.12e-19 | 1.66e-16 | 3.56e-21 |
| 3.55e-02 | 52.94 | 1.60e-05 | 8.97e-17 | 1.02e-13 | 1.71e-14 | 2.34e-19 | 3.23e-16 | 4.73e-21 |
| 3.16e-02 | 53.10 | 1.59e-05 | 1.20e-16 | 1.30e-13 | 3.08e-14 | 4.90e-19 | 6.27e-16 | 6.29e-21 |
| 2.82e-02 | 53.36 | 1.59e-05 | 1.80e-16 | 1.73e-13 | 5.08e-14 | 9.15e-19 | 1.12e-15 | 8.97e-21 |
| 2.51e-02 | 53.63 | 1.59e-05 | 2.69e-16 | 2.32e-13 | 8.38e-14 | 1.71e-18 | 2.01e-15 | 1.28e-20 |
| 2.24e-02 | 54.01 | 1.59e-05 | 4.50e-16 | 3.55e-13 | 1.21e-13 | 2.26e-18 | 2.92e-15 | 2.13e-20 |
| 1.99e-02 | 54.39 | 1.59e-05 | 7.51e-16 | 5.43e-13 | 1.75e-13 | 2.99e-18 | 4.25e-15 | 3.55e-20 |
| 1.78e-02 | 55.14 | 1.59e-05 | 1.90e-15 | 1.08e-12 | 1.93e-13 | 3.96e-18 | 6.17e-15 | 8.26e-20 |
| 1.58e-02 | 55.88 | 1.59e-05 | 4.81e-15 | 2.17e-12 | 2.13e-13 | 5.25e-18 | 8.97e-15 | 1.92e-19 |
| 1.41e-02 | 56.82 | 1.59e-05 | 1.41e-14 | 4.94e-12 | 2.32e-13 | 5.55e-18 | 1.19e-14 | 5.22e-19 |
| 1.26e-02 | 57.76 | 1.59e-05 | 4.16e-14 | 1.13e-11 | 2.53e-13 | 5.86e-18 | 1.57e-14 | 1.42e-18 |
| 1.12e-02 | 58.79 | 1.58e-05 | 1.28e-13 | 2.64e-11 | 3.14e-13 | 6.28e-18 | 1.95e-14 | 4.04e-18 |
| 1.00e-02 | 59.83 | 1.58e-05 | 3.91e-13 | 6.21e-11 | 3.89e-13 | 6.73e-18 | 2.42e-14 | 1.15e-17 |
| 8.91e-03 | 60.97 | 1.58e-05 | 1.27e-12 | 1.45e-10 | 5.16e-13 | 7.15e-18 | 2.84e-14 | 3.49e-17 |
| 7.94e-03 | 62.12 | 1.58e-05 | 4.13e-12 | 3.37e-10 | 6.85e-13 | 7.60e-18 | 3.33e-14 | 1.06e-16 |
| 7.08e-03 | 63.44 | 1.57e-05 | 1.42e-11 | 6.74e-10 | 9.22e-13 | 5.35e-18 | 2.28e-14 | 3.41e-16 |
| 6.31e-03 | 64.76 | 1.57e-05 | 4.88e-11 | 1.35e-09 | 1.24e-12 | 3.76e-18 | 1.56e-14 | 1.10e-15 |
| 5.62e-03 | 66.33 | 1.57e-05 | 1.64e-10 | 2.02e-09 | 1.64e-12 | 1.32e-17 | 8.72e-15 | 3.57e-15 |
| 5.01e-03 | 67.89 | 1.56e-05 | 5.51e-10 | 3.03e-09 | 2.17e-12 | 4.66e-17 | 4.87e-15 | 1.16e-14 |
| 4.47e-03 | 69.71 | 1.55e-05 | 1.62e-09 | 3.77e-09 | 2.85e-12 | 3.02e-16 | 4.97e-15 | 4.13e-14 |
| 3.98e-03 | 71.53 | 1.55e-05 | 4.76e-09 | 4.67e-09 | 3.74e-12 | 1.96e-15 | 5.08e-15 | 1.47e-13 |
| 3.55e-03 | 73.40 | 1.54e-05 | 7.78e-09 | 5.49e-09 | 4.87e-12 | 4.07e-15 | 6.27e-15 | 4.06e-13 |
| 3.16e-03 | 75.27 | 1.53e-05 | 1.27e-08 | 6.45e-09 | 6.33e-12 | 8.46e-15 | 7.73e-15 | 1.12e-12 |
| 2.82e-03 | 77.10 | 1.52e-05 | 1.58e-08 | 7.53e-09 | 8.19e-12 | 1.14e-14 | 9.26e-15 | 1.99e-12 |
| 2.51e-03 | 78.94 | 1.51e-05 | 1.96e-08 | 8.78e-09 | 1.06e-11 | 1.54e-14 | 1.11e-14 | 3.54e-12 |
| 2.24e-03 | 80.71 | 1.49e-05 | 2.31e-08 | 1.03e-08 | 1.37e-11 | 2.94e-14 | 1.31e-14 | 4.58e-12 |
| 1.99e-03 | 82.49 | 1.48e-05 | 2.71e-08 | 1.20e-08 | 1.78e-11 | 5.62e-14 | 1.55e-14 | 5.93e-12 |
| 1.78e-03 | 84.18 | 1.46e-05 | 3.16e-08 | 1.39e-08 | 2.30e-11 | 1.21e-13 | 1.85e-14 | 7.04e-12 |
| 1.59e-03 | 85.88 | 1.44e-05 | 3.68e-08 | 1.62e-08 | 2.97e-11 | 2.62e-13 | 2.20e-14 | 8.36e-12 |
| 1.41e-03 | 87.48 | 1.42e-05 | 4.27e-08 | 1.88e-08 | 3.83e-11 | 6.73e-13 | 2.68e-14 | 9.81e-12 |
| 1.26e-03 | 89.09 | 1.39e-05 | 4.95e-08 | 2.18e-08 | 4.95e-11 | 1.73e-12 | 3.27e-14 | 1.15e-11 |
| 1.12e-03 | 90.61 | 1.36e-05 | 5.71e-08 | 2.51e-08 | 6.38e-11 | 3.89e-12 | 4.17e-14 | 1.33e-11 |
| 1.00e-03 | 92.14 | 1.32e-05 | 6.58e-08 | 2.90e-08 | 8.23e-11 | 8.76e-12 | 5.32e-14 | 1.54e-11 |
| 8.91e-04 | 93.60 | 1.28e-05 | 7.54e-08 | 3.31e-08 | 1.05e-10 | 1.53e-11 | 7.11e-14 | 1.78e-11 |
| 7.94e-04 | 95.07 | 1.24e-05 | 8.63e-08 | 3.79e-08 | 1.35e-10 | 2.66e-11 | 9.50e-14 | 2.05e-11 |

Appendix A, Continued.

| P(bar) | T (K) | mole fraction | | | | | | |
|---|---|---|---|---|---|---|---|---|
| | | $CH_4$ | $C_2H_2$ | $C_2H_6$ | $CH_3C_2H$ | $C_4H_2$ | $CH_3$ | $CO_2$ |
| 7.08e-04 | 96.50 | 1.18e-05 | 9.79e-08 | 4.30e-08 | 1.69e-10 | 4.05e-11 | 1.30e-13 | 2.34e-11 |
| 6.31e-04 | 97.93 | 1.13e-05 | 1.11e-07 | 4.88e-08 | 2.12e-10 | 6.16e-11 | 1.77e-13 | 2.67e-11 |
| 5.62e-04 | 99.35 | 1.07e-05 | 1.25e-07 | 5.50e-08 | 2.54e-10 | 8.38e-11 | 2.41e-13 | 3.02e-11 |
| 5.01e-04 | 100.77 | 1.01e-05 | 1.40e-07 | 6.20e-08 | 3.04e-10 | 1.14e-10 | 3.27e-13 | 3.41e-11 |
| 4.47e-04 | 102.21 | 9.35e-06 | 1.54e-07 | 6.93e-08 | 3.39e-10 | 1.22e-10 | 4.41e-13 | 3.80e-11 |
| 3.98e-04 | 103.65 | 8.66e-06 | 1.70e-07 | 7.75e-08 | 3.79e-10 | 1.30e-10 | 5.94e-13 | 4.24e-11 |
| 3.55e-04 | 105.12 | 7.82e-06 | 1.84e-07 | 8.60e-08 | 3.95e-10 | 1.20e-10 | 9.66e-13 | 4.66e-11 |
| 3.16e-04 | 106.60 | 7.06e-06 | 2.00e-07 | 9.54e-08 | 4.12e-10 | 1.11e-10 | 1.57e-12 | 5.12e-11 |
| 2.82e-04 | 108.11 | 6.16e-06 | 2.13e-07 | 1.05e-07 | 4.04e-10 | 9.80e-11 | 3.66e-12 | 5.53e-11 |
| 2.51e-04 | 109.62 | 5.37e-06 | 2.27e-07 | 1.16e-07 | 3.97e-10 | 8.65e-11 | 8.51e-12 | 5.98e-11 |
| 2.24e-04 | 111.17 | 4.44e-06 | 2.35e-07 | 1.23e-07 | 3.61e-10 | 6.96e-11 | 1.48e-11 | 6.36e-11 |
| 1.99e-04 | 112.71 | 3.68e-06 | 2.44e-07 | 1.30e-07 | 3.28e-10 | 5.60e-11 | 2.56e-11 | 6.76e-11 |
| 1.78e-04 | 114.29 | 2.88e-06 | 2.39e-07 | 1.21e-07 | 2.70e-10 | 4.36e-11 | 3.33e-11 | 7.10e-11 |
| 1.59e-04 | 115.88 | 2.25e-06 | 2.35e-07 | 1.13e-07 | 2.23e-10 | 3.39e-11 | 4.34e-11 | 7.45e-11 |
| 1.41e-04 | 117.48 | 1.66e-06 | 2.10e-07 | 8.96e-08 | 1.60e-10 | 2.39e-11 | 4.71e-11 | 7.72e-11 |
| 1.26e-04 | 119.09 | 1.23e-06 | 1.88e-07 | 7.09e-08 | 1.15e-10 | 1.68e-11 | 5.11e-11 | 8.00e-11 |
| 1.12e-04 | 120.71 | 8.65e-07 | 1.47e-07 | 4.93e-08 | 6.94e-11 | 9.58e-12 | 4.93e-11 | 8.20e-11 |
| 1.00e-04 | 122.34 | 6.08e-07 | 1.15e-07 | 3.43e-08 | 4.19e-11 | 5.46e-12 | 4.76e-11 | 8.41e-11 |
| 8.91e-05 | 123.97 | 4.07e-07 | 7.97e-08 | 2.24e-08 | 2.18e-11 | 2.55e-12 | 4.35e-11 | 8.58e-11 |
| 7.94e-05 | 125.61 | 2.73e-07 | 5.52e-08 | 1.46e-08 | 1.13e-11 | 1.19e-12 | 3.98e-11 | 8.76e-11 |
| 7.08e-05 | 127.25 | 1.75e-07 | 3.40e-08 | 8.92e-09 | 5.18e-12 | 4.52e-13 | 3.48e-11 | 8.89e-11 |
| 6.31e-05 | 128.89 | 1.12e-07 | 2.09e-08 | 5.44e-09 | 2.37e-12 | 1.72e-13 | 3.04e-11 | 9.03e-11 |
| 5.62e-05 | 130.54 | 6.86e-08 | 1.12e-08 | 2.75e-09 | 9.54e-13 | 4.69e-14 | 2.37e-11 | 9.01e-11 |
| 5.01e-05 | 132.18 | 4.20e-08 | 5.99e-09 | 1.39e-09 | 3.84e-13 | 1.28e-14 | 1.85e-11 | 8.99e-11 |
| 4.47e-05 | 133.83 | 2.41e-08 | 2.82e-09 | 5.21e-10 | 1.40e-13 | 2.40e-15 | 1.17e-11 | 8.89e-11 |
| 3.98e-05 | 135.48 | 1.38e-08 | 1.33e-09 | 1.96e-10 | 5.11e-14 | 4.51e-16 | 7.38e-12 | 8.79e-11 |
| 3.55e-05 | 137.13 | 6.76e-09 | 5.58e-10 | 5.29e-11 | 1.74e-14 | 6.32e-17 | 3.27e-12 | 8.69e-11 |
| 3.16e-05 | 138.78 | 3.31e-09 | 2.34e-10 | 1.43e-11 | 5.90e-15 | 8.86e-18 | 1.45e-12 | 8.60e-11 |
| 2.82e-05 | 140.43 | 1.35e-09 | 8.87e-11 | 3.21e-12 | 1.88e-15 | 1.02e-18 | 4.33e-13 | 8.45e-11 |
| 2.51e-05 | 142.08 | 5.50e-10 | 3.36e-11 | 7.17e-13 | 6.00e-16 | 1.17e-19 | 1.29e-13 | 8.30e-11 |
| 2.24e-05 | 143.73 | 2.30e-10 | 1.18e-11 | 1.55e-13 | 1.90e-16 | 1.19e-20 | 5.40e-14 | 8.13e-11 |
| 2.00e-05 | 145.38 | 9.59e-11 | 4.17e-12 | 3.37e-14 | 6.01e-17 | 1.20e-21 | 2.26e-14 | 7.97e-11 |
| 1.78e-05 | 147.03 | 4.89e-11 | 1.39e-12 | 7.23e-15 | 1.93e-17 | 1.10e-22 | 1.39e-14 | 7.86e-11 |
| 1.58e-05 | 148.68 | 2.49e-11 | 4.60e-13 | 1.55e-15 | 6.20e-18 | 1.01e-23 | 8.54e-15 | 7.75e-11 |
| 1.41e-05 | 150.33 | 1.98e-11 | 1.44e-13 | 3.78e-16 | 1.96e-18 | 8.33e-25 | 8.16e-15 | 7.70e-11 |
| 1.26e-05 | 151.98 | 1.58e-11 | 4.53e-14 | 9.23e-17 | 6.23e-19 | 6.87e-26 | 7.80e-15 | 7.65e-11 |
| 1.12e-05 | 153.63 | 1.57e-11 | 1.37e-14 | 4.58e-17 | 1.93e-19 | 6.07e-27 | 8.62e-15 | 7.57e-11 |
| 1.00e-05 | 155.28 | 1.56e-11 | 4.13e-15 | 2.27e-17 | 5.95e-20 | 5.36e-28 | 9.53e-15 | 7.50e-11 |
| 8.91e-06 | 156.93 | 1.59e-11 | 1.22e-15 | 1.92e-17 | 1.91e-20 | 1.08e-28 | 1.07e-14 | 7.47e-11 |
| 7.94e-06 | 158.58 | 1.62e-11 | 3.60e-16 | 1.62e-17 | 6.14e-21 | 2.19e-29 | 1.20e-14 | 7.44e-11 |
| 7.08e-06 | 160.23 | 1.61e-11 | 1.18e-16 | 1.42e-17 | 3.02e-21 | 1.82e-29 | 1.31e-14 | 7.38e-11 |
| 6.31e-06 | 161.88 | 1.60e-11 | 3.85e-17 | 1.25e-17 | 1.49e-21 | 1.52e-29 | 1.43e-14 | 7.33e-11 |
| 5.62e-06 | 163.53 | 1.54e-11 | 2.05e-17 | 1.08e-17 | 1.41e-21 | 1.84e-29 | 1.55e-14 | 7.31e-11 |

Appendix A, Continued.

| P(bar) | T (K) | mole fraction | | | | | | |
|---|---|---|---|---|---|---|---|---|
| | | $CH_4$ | $C_2H_2$ | $C_2H_6$ | $CH_3C_2H$ | $C_4H_2$ | $CH_3$ | $CO_2$ |
| 5.01e-06 | 165.18 | 1.49e-11 | 1.09e-17 | 9.35e-18 | 1.33e-21 | 2.22e-29 | 1.68e-14 | 7.29e-11 |
| 4.47e-06 | 166.83 | 1.41e-11 | 9.20e-18 | 8.02e-18 | 1.52e-21 | 2.84e-29 | 1.80e-14 | 7.27e-11 |
| 3.98e-06 | 168.48 | 1.34e-11 | 7.76e-18 | 6.87e-18 | 1.73e-21 | 3.63e-29 | 1.94e-14 | 7.25e-11 |
| 3.55e-06 | 170.13 | 1.24e-11 | 7.43e-18 | 5.83e-18 | 1.96e-21 | 4.69e-29 | 2.09e-14 | 7.19e-11 |
| 3.16e-06 | 171.78 | 1.15e-11 | 7.12e-18 | 4.95e-18 | 2.23e-21 | 6.05e-29 | 2.25e-14 | 7.12e-11 |
| 2.82e-06 | 173.43 | 1.07e-11 | 7.00e-18 | 4.24e-18 | 2.49e-21 | 7.93e-29 | 2.46e-14 | 7.12e-11 |
| 2.51e-06 | 175.08 | 9.86e-12 | 6.88e-18 | 3.63e-18 | 2.78e-21 | 1.04e-28 | 2.68e-14 | 7.12e-11 |
| 2.24e-06 | 176.73 | 9.04e-12 | 6.79e-18 | 3.08e-18 | 3.00e-21 | 1.38e-28 | 2.96e-14 | 7.09e-11 |
| 1.99e-06 | 178.38 | 8.29e-12 | 6.70e-18 | 2.62e-18 | 3.23e-21 | 1.82e-28 | 3.27e-14 | 7.06e-11 |
| 1.78e-06 | 180.03 | 7.55e-12 | 6.63e-18 | 2.19e-18 | 3.37e-21 | 2.63e-28 | 3.79e-14 | 7.03e-11 |
| 1.59e-06 | 181.68 | 6.88e-12 | 6.57e-18 | 1.83e-18 | 3.51e-21 | 3.80e-28 | 4.40e-14 | 7.01e-11 |
| 1.41e-06 | 197.46 | 6.25e-12 | 6.50e-18 | 1.43e-18 | 3.56e-21 | 6.97e-28 | 6.08e-14 | 7.03e-11 |
| 1.26e-06 | 213.24 | 5.67e-12 | 6.43e-18 | 1.12e-18 | 3.62e-21 | 1.28e-27 | 8.41e-14 | 7.06e-11 |
| 1.12e-06 | 229.04 | 5.11e-12 | 6.25e-18 | 7.23e-19 | 3.77e-21 | 3.04e-27 | 1.03e-13 | 7.10e-11 |
| 1.00e-06 | 244.84 | 4.60e-12 | 6.08e-18 | 4.67e-19 | 3.92e-21 | 7.24e-27 | 1.27e-13 | 7.14e-11 |
| 8.91e-07 | 260.64 | 3.99e-12 | 5.75e-18 | 2.08e-19 | 4.29e-21 | 1.68e-26 | 1.19e-13 | 7.18e-11 |
| 7.94e-07 | 276.44 | 3.47e-12 | 5.44e-18 | 9.28e-20 | 4.69e-21 | 3.91e-26 | 1.11e-13 | 7.22e-11 |
| 7.08e-07 | 292.24 | 2.64e-12 | 4.88e-18 | 2.51e-20 | 3.68e-21 | 6.41e-26 | 6.63e-14 | 7.22e-11 |
| 6.31e-07 | 308.04 | 2.01e-12 | 4.38e-18 | 6.78e-21 | 2.89e-21 | 1.05e-25 | 3.96e-14 | 7.23e-11 |
| 5.62e-07 | 323.84 | 1.30e-12 | 3.41e-18 | 1.35e-21 | 1.40e-21 | 9.71e-26 | 1.76e-14 | 7.21e-11 |
| 5.01e-07 | 339.64 | 8.44e-13 | 2.65e-18 | 2.70e-22 | 6.78e-22 | 8.98e-26 | 7.79e-15 | 7.19e-11 |
| 4.47e-07 | 355.44 | 4.84e-13 | 1.66e-18 | 4.53e-23 | 2.10e-22 | 4.59e-26 | 3.15e-15 | 7.11e-11 |
| 3.98e-07 | 371.24 | 2.77e-13 | 1.04e-18 | 7.59e-24 | 6.49e-23 | 2.35e-26 | 1.27e-15 | 7.03e-11 |
| 3.55e-07 | 386.54 | 1.47e-13 | 5.49e-19 | 1.10e-24 | 1.42e-23 | 7.73e-27 | 5.29e-16 | 6.94e-11 |
| 3.16e-07 | 401.84 | 7.84e-14 | 2.90e-19 | 1.61e-25 | 3.12e-24 | 2.54e-27 | 2.20e-16 | 6.85e-11 |
| 2.82e-07 | 418.14 | 4.03e-14 | 1.35e-19 | 2.15e-26 | 5.67e-25 | 6.11e-28 | 1.00e-16 | 6.70e-11 |
| 2.51e-07 | 434.44 | 2.07e-14 | 6.26e-20 | 2.87e-27 | 1.03e-25 | 1.47e-28 | 4.57e-17 | 6.56e-11 |
| 2.24e-07 | 437.87 | 1.05e-14 | 2.55e-20 | 4.18e-28 | 2.10e-26 | 2.72e-29 | 2.55e-17 | 6.34e-11 |
| 2.00e-07 | 441.30 | 5.33e-15 | 1.04e-20 | 6.09e-29 | 4.26e-27 | 5.02e-30 | 1.42e-17 | 6.12e-11 |
| 1.78e-07 | 444.72 | 2.68e-15 | 3.96e-21 | 1.08e-29 | 7.99e-28 | 7.95e-31 | 8.01e-18 | 5.87e-11 |
| 1.58e-07 | 448.15 | 1.35e-15 | 1.51e-21 | 1.93e-30 | 1.50e-28 | 1.26e-31 | 4.52e-18 | 5.63e-11 |
| 1.41e-07 | 451.58 | 6.73e-16 | 5.51e-22 | 3.92e-31 | 2.57e-29 | 1.80e-32 | 2.53e-18 | 5.34e-11 |
| 1.26e-07 | 455.01 | 3.36e-16 | 2.01e-22 | 7.97e-32 | 4.42e-30 | 2.56e-33 | 1.41e-18 | 5.07e-11 |
| 1.12e-07 | 458.44 | 1.66e-16 | 7.19e-23 | 1.75e-32 | 7.19e-31 | 3.44e-34 | 7.84e-19 | 4.80e-11 |
| 1.00e-07 | 461.87 | 8.25e-17 | 2.57e-23 | 3.86e-33 | 1.17e-31 | 4.62e-35 | 4.36e-19 | 4.54e-11 |

| P(bar) | T (K) |
|---|---|
| 8.91e-08 | 465.37 |
| 7.94e-08 | 468.87 |
| 7.08e-08 | 472.37 |
| 6.31e-08 | 475.87 |
| 5.62e-08 | 479.37 |
| 5.01e-08 | 482.87 |

Appendix A, Continued.

| P(bar) | T (K) |
|---|---|
| 4.47e-08 | 486.37 |
| 3.98e-08 | 489.87 |
| 3.55e-08 | 493.37 |
| 3.16e-08 | 496.87 |
| 2.82e-08 | 500.37 |
| 2.51e-08 | 503.87 |
| 2.24e-08 | 507.37 |
| 1.99e-08 | 510.87 |
| 1.78e-08 | 514.37 |
| 1.58e-08 | 517.87 |
| 1.41e-08 | 521.37 |
| 1.26e-08 | 524.87 |
| 1.12e-08 | 528.37 |
| 1.00e-08 | 531.87 |
| 8.91e-09 | 535.37 |
| 7.94e-09 | 538.87 |
| 7.08e-09 | 542.37 |
| 6.31e-09 | 545.87 |
| 5.62e-09 | 549.37 |
| 5.01e-09 | 552.87 |
| 4.47e-09 | 564.28 |
| 3.98e-09 | 575.70 |
| 3.55e-09 | 587.11 |
| 3.16e-09 | 598.53 |
| 2.82e-09 | 609.94 |
| 2.51e-09 | 621.36 |
| 2.24e-09 | 632.77 |
| 1.99e-09 | 644.18 |
| 1.78e-09 | 655.60 |
| 1.59e-09 | 667.01 |
| 1.41e-09 | 678.43 |
| 1.26e-09 | 689.84 |
| 1.12e-09 | 701.26 |
| 1.00e-09 | 712.67 |

**Appendix B. $C_4H_2$ vapor pressure.**

Khanna et al. (1990) developed a new infrared-spectroscopic technique to measure the reduced vapor pressures of organic ices at low temperatures, and they used it to derive the $C_4H_2$ vapor pressure over $C_4H_2$ ice at temperatures as low as 127 K. Moses et al. (2005) took the Khanna et al. (1990) low-temperature data points at face value and gave them equal weight compared to higher-temperature data from other experimental techniques (e.g., Tanneberger 1933 and unpublished capacitance-manometer data from J. E. Allen, personal communication to J. Moses, 1990) when deriving the $C_4H_2$ vapor-pressure expression used in their models. However, as is shown in Table I of Khanna et al. (1990), the new infrared technique underestimates the low-temperature vapor pressures for $CO_2$ over $CO_2$ ice by factors of ~2-4 compared with other techniques, and their values for the $C_4H_2$ vapor pressure at 127-133 K may be too low by similar factors.

For the photochemical models in this paper, we assume that the Khanna et al. (1990) low-temperature vapor-pressure data are less certain by a factor of 5 in comparison with the other vapor-pressure measurements when doing our fits to the data. For a pure linear fit in log(pressure) vs. inverse temperature space, we then derive the following expression for the vapor pressure of $C_4H_2$ over $C_4H_2$ ice:

$$\log_{10}(P) = 10.2508 - 1958.25/T,$$

where $T$ is the temperature in K, and $P$ is the vapor pressure in torr. Since linear extrapolations tend to overestimate vapor pressures when extrapolated to much lower temperatures, we also tried various other expressions, including some of the more theoretically motivated expressions described in Reid et al. (1987). Since a fit to the "modified Miller equation" (see eqn. 7-6.2 on page 215 of Reid et al. 1987) results in a virtually identical curve to a simple linear + log fit, we settle on the latter. The expression we derive for this linear + log fit in log(pressure) vs. inverse temperature space is:

$$\log_{10}(P) = 5.4094 - 3292.16/T + 16.534\log_{10}(1000/T),$$

for $T$ in K and $P$ in torr.

**Figure B-1** shows how our vapor-pressure expressions compare with available data (Tanneberger 1933; Khanna et al. 1990; J.E. Allen, personal communication, 1990) and with other theoretical expressions. The vapor-pressure expression used in Moses et al. (2005), when extrapolated to relevant Uranus temperatures (e.g., vertical dotted lines in **Fig. B-1**), results in vapor pressures many orders of magnitude lower than with the newer expressions. These low vapor pressures keep the $C_4H_2$ column abundance from becoming significant in the older Moses et al. (2005) models. The photochemical production of $C_4H_2$ then occurs largely within its condensation region, and there is insufficient production of $C_4H_2$ at altitudes above its saturation level to build up an observable column density. With the new expressions, there is a larger altitude range available for $C_4H_2$ production above its condensation level, and thus more $C_4H_2$ vapor is present. The Romani and Atreya (1989) expression for the $C_4H_2$ vapor pressure lies in between our above expressions at relevant Uranus temperatures, and goes through the Khanna et al. (1990) low-temperature points, but it underestimates the vapor pressures at higher temperatures and should not be used for $T > 150$ K.

Models that adopt the linear + log expression described above still significantly underestimate the $C_4H_2$ column abundance needed to explain the Spitzer observations. The new linear expression, on the other hand, produces a $C_4H_2$ profile in our nominal model that needs to be scaled by only 1.16 to provide the best fit to the Spitzer data. This suggests that the linear expression is reasonable. However, we encourage new measurements of the $C_4H_2$ vapor pressure over $C_4H_2$ ice at temperatures down as low as 90 K, if experimentally feasible. Such measurements will have potential implications for Saturn, Titan, and Neptune, as well.

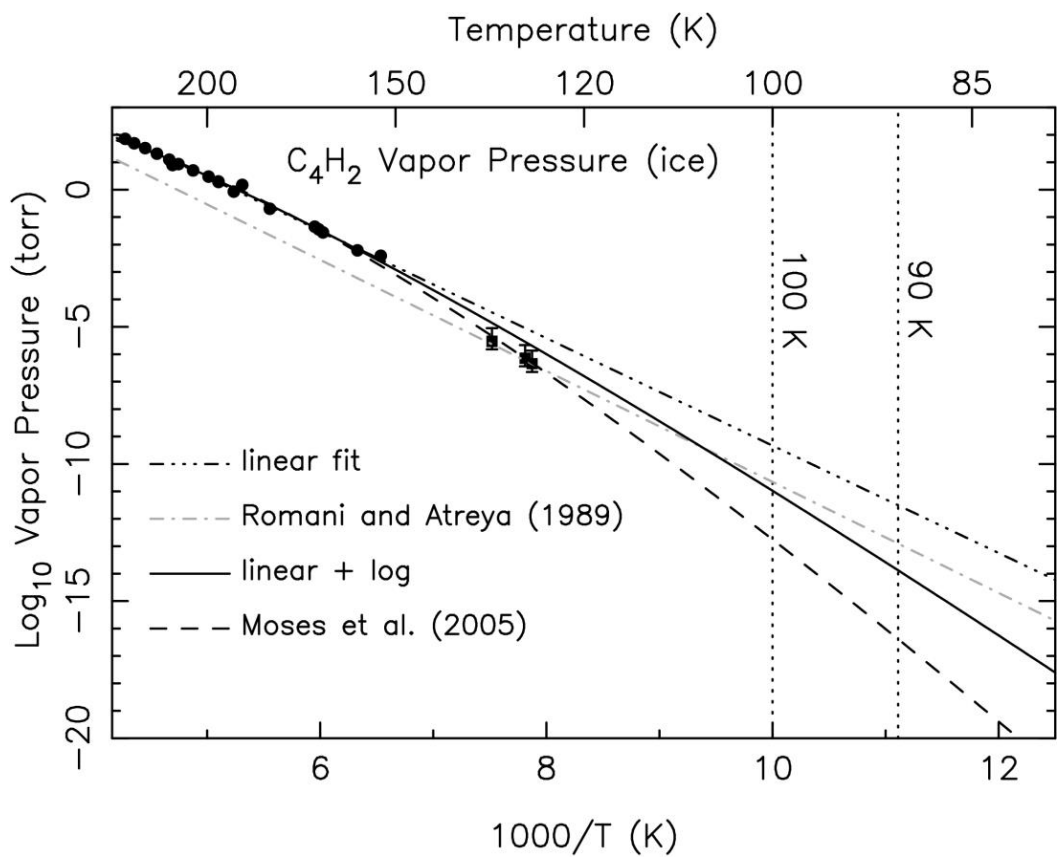

**Figure B-1**. The vapor pressure of $C_4H_2$ over condensed $C_4H_2$ from various measurements (squares; from Tanneberger 1933; Khanna et al. 1990; and unpublished capacitance-manometer data from J. E. Allen, personal communication, 1990), combined with various fits to the data. The dashed line represents the expression used by Moses et al. (2005), the gray dot-dashed line represents the expression used by Romani and Atreya (1989), and the other lines represent fits to the data for a linear fit (triple-dot-dashed line) and a linear-plus-log fit (solid line), using weighting-factors that assume the three low-temperature points from Khanna et al. (1990) are uncertain by a factor of 5 (see text for further details). The temperatures relevant to the $C_4H_2$ condensation region in the lower stratosphere of Uranus lie between the two vertical dotted lines (90 and 100 K). Note that significant uncertainties in the vapor pressures at relevant Uranian temperatures are introduced through the required extrapolation. Our photochemical-modeling results and comparisons with the Spitzer data suggest that the vapor pressures adopted by Moses et al. (2005; dashed line) are too low, and we now adopt the linear fit (triple-dot-dashed line) throughout the current analysis.